\documentclass[showpacs,preprintnumbers,amsmath,amssymb]{revtex4}

\newcommand{\bea}{\begin{eqnarray}}
\newcommand{\eea}{\end{eqnarray}}

\begin{document}
\title{Third-order cosmological perturbations of zero-pressure multi-component fluids: \\
       Pure general relativistic nonlinear effects}
\author{Jai-chan Hwang}
 \email{jchan@knu.ac.kr}
 \affiliation{Department of Astronomy and Atmospheric Sciences,
              Kyungpook National University, Taegu, Korea}
\author{Hyerim Noh}
 \email{hr@kasi.re.kr}
 \affiliation{Korea Astronomy and Space Science Institute,
              Daejon, Korea}

\date{\today}

\begin{abstract}

Present expansion stage of the universe is believed to be mainly
governed by the cosmological constant, collisionless dark matter and
baryonic matter. The latter two components are often modeled as
zero-pressure fluids. In our previous work we have shown that to the
second-order cosmological perturbations, the relativistic equations
of the zero-pressure, irrotational, multi-component fluids in a
spatially near flat background effectively coincide with the
Newtonian equations. As the Newtonian equations only have quadratic
order nonlinearity, it is practically interesting to derive the
potential third-order perturbation terms in general relativistic
treatment which correspond to pure general relativistic corrections.
In our previous work we have shown that even in a single component
fluid there exists substantial amount of pure relativistic
third-order correction terms. We have, however, shown that those
correction terms are independent of the horizon scale, and are quite
small ($\sim 5 \times 10^{-5}$ smaller compared with the
relativistic/Newtonian second-order terms) due to the weak level
anisotropy of the cosmic microwave background radiation. Here, we
present pure general relativistic correction terms appearing in the
third-order perturbations of the multi-component zero-pressure
fluids. As a result we show that, as in a single component
situation, the third-order correction terms are quite small ($\sim 5
\times 10^{-5}$ smaller) in the context of the evolution of our
universe. Still, there do exist pure general relativistic correction
terms in third-order perturbations which could potentially become
important in future development of precision cosmology. We include
the cosmological constant in all our analyses.

\end{abstract}
\noindent 

\maketitle

\tableofcontents

%
%
\section{Introduction}
                                          \label{sec:Introduction}

Recently, we have been presenting a series of work based on our
theoretical study of relativistic nonlinear cosmological
perturbations \cite{NL,second-order,second-order-SG,third-order}. We
have shown that to the second-order perturbations, general
relativistic equations of a zero-pressure, irrotational fluid in a
spatially flat background have exact Newtonian correspondence except
for the presence of the gravitational wave contributions
\cite{NL,second-order}. In an accompanying paper
\cite{second-order-multi} we have relaxed all the assumptions we
made in the second-order perturbations in \cite{NL,second-order},
and have derived pure general relativistic effects from the
pressures, rotation, spatial curvature, and multi-component. In that
work, we have shown that except for the multi-component situation,
relaxing any of the other three assumptions leads to pure general
relativistic correction terms appearing in the second order.
Pressures are intrinsically general relativistic even in the
background and the linear-order perturbations. The presence of
background curvature leads to first non-vanishing relativistic
correction terms appearing in the second order. The rotational
perturbations generally lead to relativistic correction terms which
become negligible in the small-scale (sub-horizon scale) limit, thus
having relativistic/Newtonian correspondence in that limit. In the
case of zero-pressure, irrotational multi-component fluids in a flat
background, effectively we have exact relativistic/Newtonian
correspondence even in the multi-component situation; this will be
summarized in the later section of this work.

The relativistic/Newtonian correspondence in the background world
model was known in the zero-pressure medium by the work of Friedmann
in 1922 \cite{Friedmann-1922} in the context of Einstein's gravity,
and by the work of Milne and McCrea in 1934 \cite{Milne-1934} in the
context of Newton's gravity; the latter Newtonian derivation is
later known to be a guided one by the already derived Einstein
gravity result \cite{Layzer-1954}. In the case of linear
perturbations, the relativistic/Newtonian correspondence was also
known in the zero-pressure medium by the work of Lifshitz in 1946
\cite{Lifshitz-1946} in the context of Einstein's gravity, and by
the work of Bonnor in 1957 \cite{Bonnor-1957} in the context of
Newton's gravity. The fully nonlinear perturbation equations in the
context of Newtonian cosmology in a zero-pressure medium were known
in a textbook by Peebles in 1980 \cite{Peebles-1980}. The Einstein's
gravity counterpart of the nonlinearly perturbed cosmological
medium, especially the continued relativistic/Newtonian
correspondence even to the second order, was first shown only
recently in our works in \cite{NL,second-order}.

Meanwhile, in \cite{third-order}, we derived pure general
relativistic correction terms appearing in  the third-order single
component, zero-pressure, irrotational fluid in a flat background.
Thus, now it is a natural step to find out the potential third-order
pure general relativistic correction terms appearing in the
multi-component, zero-pressure, irrotational fluids in a flat
background. As the Newtonian system has only quadratic nonlinearity
even in the multi-component situation, see Sec.\ II of
\cite{second-order-multi}, any nonvanishing third-order terms can be
regarded as pure general relativistic corrections. The situation is
also practically important because current stage of the universe is
supposed to be dominated by two zero-pressure components and the
cosmological constant.  We will include the effect of cosmological
constant in all our analyses and equations in this work which is
also true in our previous works in
\cite{NL,second-order,second-order-SG,third-order,second-order-multi}.

In Sec.\ \ref{sec:third-order-quantities} we present the metric and
fluid quantities perturbed to the third order which will be required
in our calculation. We present fluid quantities for most general
fluids with pressures and stresses which will turn out to be
important even in the zero-pressure situation in our main analysis.
As in the single component case in \cite{third-order}, under our
proper choice of variables and gauges we do not need third-order
perturbations of the connection or curvature tensor. In Sec.\
\ref{sec:second-order} we summarize the relativistic equations to
the second order, and their effective correspondence with the
Newtonian ones even in the multi-component situation. In Sec.\
\ref{sec:third-order-equations} we derive the general relativistic
third-order terms and present equations in the context of Newtonian
gravity with pure general relativistic corrections. We compare our
equations in the multi-component case with the previously derived
ones in a single component. Section \ref{sec:Discussion} is a
discussion. We often set $c \equiv 1$, but recover $c$ in the
Newtonian context presentation.

%
%
\section{Third-order perturbations}
                                        \label{sec:third-order-quantities}

\subsection{The covariant and ADM equations}
                                        \label{sec:covariant-eqs}

In the following we summarize the basic sets of covariant equations
and ADM equations we need in our analysis. These equations, except
for the covariant equations of individual component, are presented
Sec.\ II of \cite{NL}; notations can be found in that work. For
original studies of the covariant and the ADM equations, see
\cite{covariant}, and \cite{ADM}, respectively. Although, we will
use the ADM equations in our calculation, the covariant equations
show another aspects of the same fully nonlinear system of
Einstein's equation.

The energy-momentum tenor of a fluid can be decomposed into fluid
quantities as \bea
   & & \tilde T_{ab} = \tilde \mu \tilde u_a \tilde u_b
       + \tilde p \left( \tilde u_a \tilde u_b + \tilde g_{ab}
       \right)
       + \tilde q_a \tilde u_b + \tilde q_b \tilde u_a
       + \tilde \pi_{ab}.
   \label{Tab}
\eea Without losing generality, we take the energy frame, thus set
$\tilde q_a \equiv 0$. This decomposition is valid even in the
multiple component fluids; in such a case the above fluid quantities
can be regarded as collective fluid quantities. In the
multi-component case we introduce the energy-momentum tensor and
fluid quantities of individual component as \bea
   & & \tilde T_{ab} \equiv \sum_j \tilde T_{(j)ab},
   \nonumber \\
   & & \tilde T_{(i)ab} = \tilde \mu_{(i)} \tilde u_{(i)a} \tilde
       u_{(i)b}
       + \tilde p_{(i)} \left( \tilde u_{(i)a} \tilde u_{(i)b}
       + \tilde g_{ab}
       \right)
       + \tilde \pi_{(i)ab},
   \label{Tab-i}
\eea where, without losing generality, we also took the energy-frame
condition for each component, thus set $\tilde q_{(i)a} \equiv 0$.
For interactions among components we introduce \bea
   & & \tilde T^{\;\;\;\; b}_{(i)a;b} \equiv \tilde I_{(i)a}, \quad
       \sum_j \tilde I_{(j)a} = 0.
\eea

In a single component situation, taking the energy-frame (thus,
setting $\tilde q_a \equiv 0$), the energy conservation equation,
the momentum conservation equation, and the Raychaudhury equation
are \bea
   & & \tilde {\dot {\tilde \mu}}
       + \left( \tilde \mu + \tilde p \right) \tilde \theta
       + \tilde \pi^{ab} \tilde \sigma_{ab}
       = 0,
   \label{E-conserv-cov} \\
   & & \left( \tilde \mu + \tilde p \right) \tilde a_a
       + \tilde h^b_a \left( \tilde p_{,b}
       + \tilde \pi^c_{b;c} \right)
       = 0,
   \label{Mom-conserv-cov} \\
   & & \tilde {\dot {\tilde \theta}}
       + {1 \over 3} \tilde \theta^2
       - \tilde a^a_{\;\; ;a}
       + \tilde \sigma^{ab} \tilde \sigma_{ab}
       - \tilde \omega^{ab} \tilde \omega_{ab}
       + 4 \pi G \left( \tilde \mu + 3 \tilde p \right)
       - \Lambda
       = 0.
   \label{Raychaudhury-eq-cov}
\eea In the multi-component case, taking the energy-frame for
individual component (thus, setting $\tilde q_{(i)a} \equiv 0$) we
have \bea
   & & \tilde {\dot {\tilde \mu}}_{(i)}
       + \left( \tilde \mu_{(i)} + \tilde p_{(i)} \right)
       \tilde \theta_{(i)}
       + \tilde \pi^{ab}_{(i)} \tilde \sigma_{(i)ab}
       = - \tilde u^a_{(i)} \tilde I_{(i)a},
   \label{E-conserv-cov-i} \\
   & & \left( \tilde \mu_{(i)} + \tilde p_{(i)} \right)
       \tilde a_{(i)a}
       + \tilde h_{(i)a}^{\;\;\;\;b}
       \left( \tilde p_{(i),b}
       + \tilde \pi_{(i)b;c}^{\;\;\;\;c} \right)
       = \tilde h_{(i)a}^{\;\;\;\;b} \tilde I_{(i)b},
   \label{Mom-conserv-cov-i} \\
   & & \tilde {\dot {\tilde \theta}}_{(i)}
       + {1 \over 3} \tilde \theta^2_{(i)}
       - \tilde a^a_{(i);a}
       + \tilde \sigma^{ab}_{(i)} \tilde \sigma_{(i)ab}
       - \tilde \omega^{ab}_{(i)} \tilde \omega_{(i)ab}
       - 4 \pi G \left( \tilde \mu - 3 \tilde p
       - 2 \tilde T_{ab} \tilde u^a_{(i)} \tilde u^b_{(i)} \right)
       - \Lambda
       = 0.
   \label{Raychaudhury-eq-cov-i}
\eea In order to handle cosmological perturbations to the third
order, we also need the momentum constraint equation for a
collective fluid \bea
   & & \tilde h_{ab} \left(
       \tilde \omega^{bc}_{\;\;\; ;c}
       - \tilde \sigma^{bc}_{\;\;\; ;c}
       + {2 \over 3} \tilde \theta^{,b} \right)
       + \left( \tilde \omega_{ab}
       + \tilde \sigma_{ab} \right) \tilde a^b = 0.
   \label{Mom-constraint-cov}
\eea

In the ADM formulation, the energy conservation, momentum
conservation, and trace of ADM propagation equations are \bea
   & & E_{,0} N^{-1} - E_{,\alpha} N^\alpha N^{-1}
       - K \left( E + {1 \over 3} S \right)
       - \bar S^{\alpha\beta} \bar K_{\alpha\beta}
       + N^{-2} \left( N^2 J^\alpha \right)_{:\alpha}
       = 0,
   \label{E-conservation-ADM} \\
   & & J_{\alpha,0} N^{-1} - J_{\alpha:\beta} N^\beta N^{-1}
       - J_\beta N^\beta_{\;\;\;:\alpha} N^{-1} - K J_\alpha
       + E N_{,\alpha} N^{-1}
       + S^\beta_{\alpha:\beta}
       + S_\alpha^\beta N_{,\beta} N^{-1} = 0,
   \label{Mom-conservation-ADM} \\
   & & K_{,0} N^{-1} - K_{,\alpha} N^\alpha N^{-1}
       + N^{:\alpha}_{\;\;\;\;\alpha} N^{-1}
       - \bar K^{\alpha\beta} \bar K_{\alpha\beta}
       - {1\over 3} K^2 - 4 \pi G \left( E + S \right) + \Lambda = 0.
   \label{Trace-prop-ADM}
\eea In the multi-component case we have \bea
   & & E_{(i),0} N^{-1} - E_{(i),\alpha} N^\alpha N^{-1}
       - K \left( E_{(i)} + {1 \over 3} S_{(i)} \right)
       - \bar S^{\alpha\beta}_{(i)} \bar K_{\alpha\beta}
       + N^{-2} \left( N^2 J^\alpha_{(i)} \right)_{:\alpha}
       = - {1 \over N} \left( \tilde I_{(i)0}
       - \tilde I_{(i)\alpha} N^\alpha \right),
   \label{E-conservation-ADM-i} \\
   & & J_{(i)\alpha,0} N^{-1} - J_{(i)\alpha:\beta} N^\beta N^{-1}
       - J_{(i)\beta} N^\beta_{\;\;\;:\alpha} N^{-1} - K J_{(i)\alpha}
       + E_{(i)} N_{,\alpha} N^{-1}
       + S^{\;\;\;\;\beta}_{(i)\alpha:\beta}
       + S_{(i)\alpha}^{\;\;\;\;\beta} N_{,\beta} N^{-1}
       = \tilde I_{(i)\alpha}.
   \label{Mom-conservation-ADM-i}
\eea The momentum constraint equation is \bea
   & & \bar K^\beta_{\alpha:\beta}
       - {2 \over 3} K_{,\alpha}
       = 8 \pi G J_\alpha.
   \label{Mom-constraint-ADM}
\eea

The above sets of equations are only parts of the covariant and the
ADM equations; for complete sets, see
\cite{covariant,ADM,Bardeen-1980,HV-1990,NL}. We will show that, by
taking proper choice of gauges, the scalar-type perturbations to the
third order can be derived from either of the above sets of
equations. We will present the derivation based on the ADM
equations, because the covariant formalism often requires lengthier
calculation in our particular case; of course, the covariant
equations also give the same result. As we will show, however, we
use both formulations simultaneously depending on the convenience.

\subsection{Metric}

Our metric convention is the same as in \cite{NL} \bea
   & & \tilde g_{00} = - a^2 \left( 1 + 2 A \right), \quad
       \tilde g_{0\alpha} = - a^2 B_\alpha, \quad
       \tilde g_{\alpha\beta}
       = a^2 \left( g^{(3)}_{\alpha\beta}
       + 2 C_{\alpha\beta} \right),
\eea where tensor indices of $B_\alpha$ and $C_{\alpha\beta}$ are
based on $g^{(3)}_{\alpha\beta}$. To the third order, the inverse
metric becomes \bea
   & & \tilde g^{00} = - {1 \over a^2} \left(
       1
       - 2 A
       + 4 A^2
       - B^\alpha B_\alpha
       - 8 A^3
       + 4 A B^\alpha B_\alpha
       + 2 B^\alpha B^\beta C_{\alpha\beta} \right),
   \nonumber \\
   & & \tilde g^{0\alpha} = - {1 \over a^2} \left[
       B^\alpha
       - 2 A B^\alpha
       - 2 B^\beta C^\alpha_\beta
       + B^\alpha \left( 4 A^2 - B^\beta B_\beta \right)
       + 4 C^\alpha_\beta \left( A B^\beta
       + B^\gamma C^\beta_\gamma \right) \right],
   \nonumber \\
   & & \tilde g^{\alpha\beta} = {1 \over a^2} \left(
       g^{(3)\alpha\beta}
       - 2 C^{\alpha\beta}
       - B^\alpha B^\beta
       + 4 C^\alpha_\gamma C^{\beta\gamma}
       + 2 A B^\alpha B^\beta
       + 2 B^\alpha B^\gamma C^\beta_\gamma
       + 2 B^\beta B^\gamma C^\alpha_\gamma
       - 8 C^\alpha_\gamma C^\beta_\delta C^{\gamma\delta} \right).
\eea In order to derive perturbation equations to the third order,
we need the connection only to the second-order. These are presented
in Eq.\ (52) of \cite{NL}.

The ADM metric variables follow from Eq.\ (2) of \cite{NL} as \bea
   & & N = a \left( 1 + A
       - {1 \over 2} A^2 + {1 \over 2} B^\alpha B_\alpha
       + {1 \over 2} A^3
       - {1 \over 2} A B^\alpha B_\alpha
       - B^\alpha B^\beta C_{\alpha\beta} \right),
   \nonumber \\
   & & N_\alpha = - a^2 B_\alpha, \quad
       N^\alpha = - B^\alpha
       + 2 B^\beta C^\alpha_\beta
       - 4 B^\beta C^\alpha_\gamma C^\gamma_\beta,
   \nonumber \\
   & & h_{\alpha\beta} = a^2 \left( g^{(3)}_{\alpha\beta}
       + 2 C_{\alpha\beta} \right), \quad
       h^{\alpha\beta} = {1 \over a^2} \left(
       g^{(3)\alpha\beta}
       - 2 C^{\alpha\beta}
       + 4 C^\alpha_\gamma C^{\beta\gamma}
       - 8 C^\alpha_\gamma C^\beta_\delta C^{\gamma\delta} \right),
\eea where tensor index of $N_\alpha$ is based on $h_{\alpha\beta}$
as the metric; $h^{\alpha\beta}$ is an inverse metric of
$h_{\alpha\beta}$ .

\subsection{Fluid quantities}

We introduce perturbations of fluid quantities as \bea
   & & \tilde \mu = \mu + \delta \mu, \quad
       \tilde p = p + \delta p, \quad
       \tilde u_\alpha \equiv a v_\alpha, \quad
       \tilde \pi_{\alpha\beta} \equiv a^2 \Pi_{\alpha\beta},
   \label{perturbed-fluid-quantities}
\eea where tensor indices of $v_\alpha$ and $\Pi_{\alpha\beta}$ are
based on $g^{(3)}_{\alpha\beta}$. The above fluid quantities can be
regarded as collective fluid quantities in the case of
multi-component fluids, see below Eq.\ (\ref{Tab-pert}). Although we
will consider zero-pressure fluids, it is important to keep the
perturbed pressure ($\delta p$) and anisotropic stress
($\Pi_{\alpha\beta}$) because the collective pressure and
anisotropic stress do not vanish to nonlinear order in the
multi-component fluids even in the zero-pressure case, see Eq.\
(\ref{total-fluid-quantities-p=0}). In any case, for later
convenience, in this section we will present fluid quantities for
most general fluids.

Components of the four-vector $\tilde u_a$ are \bea
   & & \tilde u_\alpha \equiv a v_\alpha,
   \nonumber \\
   & & \tilde u_0 = - a \Big[
       1 + A - {1 \over 2} A^2
       + {1 \over 2} \left( v^\alpha + B^\alpha \right) \left( v_\alpha +
       B_\alpha \right)
       + {1 \over 2} A^3
       + {1 \over 2} A \left( v^\alpha v_\alpha
       - B^\alpha B_\alpha \right)
       - C_{\alpha\beta} \left( v^\alpha + B^\alpha \right)
       \left( v^\beta + B^\beta \right) \Big],
   \nonumber \\
   & & \tilde u^\alpha = {1 \over a} \Big[
       v^\alpha + B^\alpha - A B^\alpha
       - 2 C^{\alpha\beta} \left( v_\beta + B_\beta \right)
       + {3 \over 2} A^2 B^\alpha
       + 2 A B^\beta C^\alpha_\beta
       + {1 \over 2} B^\alpha \left( v^\beta v_\beta
       - B^\beta B_\beta \right)
       + 4 C^\alpha_\beta C^\beta_\gamma \left( v^\gamma + B^\gamma
       \right) \Big],
   \nonumber \\
   & & \tilde u^0 = {1 \over a} \left[
       1 - A + {3 \over 2} A^2
       + {1 \over 2} \left( v^\alpha v_\alpha - B^\alpha B_\alpha
       \right)
       - {5 \over 2} A^3
       - {1 \over 2} A \left( v^\alpha v_\alpha
       - 3 B^\alpha B_\alpha \right)
       - C_{\alpha\beta} \left( v^\alpha v^\beta
       - B^\alpha B^\beta \right) \right].
   \label{u_a-v}
\eea In \cite{NL}, instead of $v_\alpha$, we used $V_\alpha$ defined
as \bea
   & & \tilde u^\alpha \equiv {1 \over a} V^\alpha.
   \label{u_a-V}
\eea Thus, we have \bea
   & & v_\alpha = V_\alpha - B_\alpha
       + A B_\alpha + 2 C_{\alpha\beta} V^\beta
       - {3 \over 2} A^2 B_\alpha
       - B_\alpha V_\beta \left( {1 \over 2} V^\beta - B^\beta
       \right).
\eea Components of $\tilde \pi_{ab}$ are \bea
   & & \tilde \pi_{\alpha\beta} \equiv a^2 \Pi_{\alpha\beta}, \quad
       \tilde \pi_{0\alpha}
       = - a^2 \Pi_{\alpha\beta} \left[ v^\beta + B^\beta
       + A v^\beta
       - 2 C^{\beta\gamma} \left( v_\gamma + B_\gamma \right)
       \right], \quad
       \tilde \pi_{00}
       = a^2 \Pi_{\alpha\beta} \left( v^\alpha + B^\alpha \right)
       \left( v^\beta + B^\beta \right).
\eea From $\tilde \pi^c_c \equiv 0$ we have \bea
   & & \Pi^\alpha_\alpha
       - 2 C^{\alpha\beta} \Pi_{\alpha\beta}
       + \left( 4 C^\alpha_\gamma C^{\beta\gamma}
       - v^\alpha v^\beta \right) \Pi_{\alpha\beta} = 0.
\eea To the third order, the energy-momentum tensor becomes \bea
   & & \tilde T^0_0
       = - \mu - \delta \mu
       - \left( \mu + p \right) v^\alpha \left( v_\alpha + B_\alpha
       \right)
       + \left( \mu + p \right) \left[ A B_\alpha
       + 2 C_{\alpha\beta} \left( v^\beta + B^\beta \right) \right]
       v^\alpha
   \nonumber \\
   & & \qquad
       - \left( \delta \mu + \delta p \right) v^\alpha \left(
       v_\alpha + B_\alpha \right)
       - \Pi_{\alpha\beta} v^\alpha \left( v^\beta + B^\beta \right),
   \nonumber \\
   & & \tilde T^0_\alpha
       = \left( \mu + p \right) v_\alpha
       - \left( \mu + p \right) A v_\alpha
       + \left( \delta \mu + \delta p \right) v_\alpha
       + \Pi_{\alpha\beta} v^\beta
       + {1 \over 2} \left( \mu + p \right)
       \left( 3 A^2 + v^\beta v_\beta - B^\beta B_\beta \right)
       v_\alpha
   \nonumber \\
   & & \qquad
       - \left( \delta \mu + \delta p \right) A v_\alpha
       - \Pi_{\alpha\beta} \left( A v^\beta
       + 2 C^{\beta\gamma} v_\gamma \right),
   \nonumber \\
   & & \tilde T^\alpha_\beta
       = p \delta^\alpha_\beta
       + \delta p \delta^\alpha_\beta
       + \Pi^\alpha_\beta
       + \left( \mu + p \right) \left( v^\alpha + B^\alpha \right)
       v_\beta
       - 2 C^{\alpha\gamma} \Pi_{\beta\gamma}
       - \left( \mu + p \right) \left[ A B^\alpha
       + 2 C^{\alpha\gamma} \left( v_\gamma + B_\gamma \right)
       \right] v_\beta
   \nonumber \\
   & & \qquad
       + \left( \delta \mu + \delta p \right) \left( v^\alpha
       + B^\alpha \right) v_\beta
       + \Pi_{\beta\gamma} \left( B^\alpha v^\gamma
       + 4 C^\alpha_\delta C^{\gamma\delta} \right).
   \label{Tab-pert}
\eea

Above relations of four-vectors, energy-momentum tensor, and fluid
quantities are derived for a single component fluid. However, these
are also valid as the collective component in the case of
multi-component fluids. We can easily see that, by replacing these
quantities with the ones of individual component, the same relations
are valid for individual component as well. That is, for an
individual component, say $i$-th component, by replacing \bea
   & & \tilde T_{ab}, \quad
       \tilde \mu, \quad
       \tilde p, \quad
       \tilde u_a, \quad
       \tilde \pi_{ab}, \quad
       \mu, \quad
       p, \quad
       \delta \mu, \quad
       \delta p, \quad
       V_\alpha, \quad
       v_\alpha, \quad
       \Pi_{\alpha\beta},
   \label{total}
\eea with \bea
   & & \tilde T_{(i)ab}, \quad
       \tilde \mu_{(i)}, \quad
       \tilde p_{(i)}, \quad
       \tilde u_{(i)a}, \quad
       \tilde \pi_{(i)ab}, \quad
       \mu_{(i)}, \quad
       p_{(i)}, \quad
       \delta \mu_{(i)}, \quad
       \delta p_{(i)}, \quad
       V_{(i)\alpha}, \quad
       v_{(i)\alpha}, \quad
       \Pi_{(i)\alpha\beta},
   \label{individual}
\eea Eqs.\ (\ref{perturbed-fluid-quantities})-(\ref{Tab-pert}) are
valid for the $i$-th component.

From Eqs.\ (\ref{Tab}),(\ref{Tab-pert}) we can express the
collective fluid quantities in terms of the individual one. To the
background order we have \bea
   & & \mu = \sum_j \mu_{(j)}, \quad
       p = \sum_j p_{(j)}.
\eea To the third order in perturbations, we can show \bea
   & & \delta \mu
       + \left[ \left( \mu + p \right) v^\alpha
       + \left( \delta \mu + \delta p \right) v^\alpha
       + \Pi^\alpha_\beta v^\beta \right] v_\alpha
       - 2 \left( \mu + p \right) C_{\alpha\beta} v^\alpha v^\beta
   \nonumber \\
   & & \qquad
       = \sum_j \Big\{
       \delta \mu_{(j)}
       + \left[ \left( \mu_{(j)} + p_{(j)} \right) v_{(j)}^\alpha
       + \left( \delta \mu_{(j)} + \delta p_{(j)} \right) v_{(j)}^\alpha
       + \Pi^{\;\;\;\;\alpha}_{(j)\beta} v_{(j)}^\beta \right] v_{(j)\alpha}
       - 2 \left( \mu_{(j)} + p_{(j)} \right)
       C_{\alpha\beta} v_{(j)}^\alpha v_{(j)}^\beta
       \Big\},
   \nonumber \\
   & & \delta p
       + {1 \over 3}
       \left[ \left( \mu + p \right) v^\alpha
       + \left( \delta \mu + \delta p \right) v^\alpha
       + \Pi^\alpha_\beta v^\beta \right] v_\alpha
       - {2 \over 3} \left( \mu + p \right) C_{\alpha\beta} v^\alpha v^\beta
   \nonumber \\
   & & \qquad
       = \sum_j \Big\{
       \delta p_{(j)}
       + {1 \over 3}
       \left[ \left( \mu_{(j)} + p_{(j)} \right) v_{(j)}^\alpha
       + \left( \delta \mu_{(j)} + \delta p_{(j)} \right) v_{(j)}^\alpha
       + \Pi^{\;\;\;\;\alpha}_{(j)\beta} v_{(j)}^\beta \right] v_{(j)\alpha}
       - {2 \over 3} \left( \mu_{(j)} + p_{(j)} \right)
       C_{\alpha\beta} v_{(j)}^\alpha v_{(j)}^\beta
       \Big\},
   \nonumber \\
   & & \left( \mu + p \right) v_\alpha
       + \left[ \delta \mu + \delta p
       + {4 \over 3}  \left( \mu + p \right) v^\beta v_\beta
       \right] v_\alpha
       + \left[ \Pi^\beta_\alpha
       + \left( \mu + p \right)
       \left( v_\alpha v^\beta
       - {1 \over 3} \delta^\beta_\alpha v^\gamma v_\gamma \right)
       \right] v_\beta
       - {3 \over 2} \left( \mu + p \right) v_\alpha
       v^\beta v_\beta
   \nonumber \\
   & & \qquad
       - 2 \Pi_{\alpha\beta} C^{\beta\gamma} v_\gamma
       = \sum_j \Bigg\{\left( \mu_{(j)} + p_{(j)} \right)
       v_{(j)\alpha}
       + \left[ \delta \mu_{(j)} + \delta p_{(j)}
       + {4 \over 3}  \left( \mu_{(j)} + p_{(j)} \right) v_{(j)}^\beta
       v_{(j)\beta}
       \right] v_{(j)\alpha}
   \nonumber \\
   & & \qquad
       + \left[ \Pi^{\;\;\;\;\beta}_{(j)\alpha}
       + \left( \mu_{(j)} + p_{(j)} \right)
       \left( v_{(j)\alpha} v_{(j)}^\beta
       - {1 \over 3} \delta^\beta_\alpha v_{(j)}^\gamma v_{(j)\gamma} \right)
       \right] v_{(j)\beta}
       - {3 \over 2} \left( \mu_{(j)} + p_{(j)} \right)
       v_{(j)\alpha}
       v_{(j)}^\beta v_{(j)\beta}
       - 2 \Pi_{(j)\alpha\beta} C^{\beta\gamma} v_{(j)\gamma}
       \Bigg\},
   \nonumber \\
   & & \Pi^\alpha_\beta
       + \left[ \left( \mu + p \right) v^\alpha
       + \left( \delta \mu + \delta p \right) v^\alpha
       + \Pi^\alpha_\gamma v^\gamma \right] v_\beta
       - {2 \over 3} \left( \mu + p \right) C^\alpha_\beta
       v^\gamma v_\gamma
       - \Pi^{\alpha\gamma} v_\beta v_\gamma
   \nonumber \\
   & & \qquad
       - {1 \over 3} \delta^\alpha_\beta
       \left\{ \left[ \left( \mu + p \right) v^\gamma
       + \left( \delta \mu + \delta p \right) v^\gamma
       + \Pi^\gamma_\delta v^\delta \right] v_\gamma
       - 2 \left( \mu + p \right) C_{\gamma\delta} v^\gamma v^\delta
       \right\}
   \nonumber \\
   & & \qquad
       = \sum_j \Bigg\{ \Pi^{\;\;\;\;\alpha}_{(j)\beta}
       + \left[ \left( \mu_{(j)} + p_{(j)} \right) v_{(j)}^\alpha
       + \left( \delta \mu_{(j)} + \delta p_{(j)} \right) v_{(j)}^\alpha
       + \Pi^{\;\;\;\;\alpha}_{(j)\gamma} v_{(j)}^\gamma \right] v_{(j)\beta}
   \nonumber \\
   & & \qquad
       - {2 \over 3} \left( \mu_{(j)} + p_{(j)} \right) C^\alpha_\beta
       v_{(j)}^\gamma v_{(j)\gamma}
       - \Pi_{(j)}^{\alpha\gamma} v_{(j)\beta} v_{(j)\gamma}
   \nonumber \\
   & & \qquad
       - {1 \over 3} \delta^\alpha_\beta
       \left\{ \left[ \left( \mu_{(j)} + p_{(j)} \right) v_{(j)}^\gamma
       + \left( \delta \mu_{(j)} + \delta p_{(j)} \right) v_{(j)}^\gamma
       + \Pi^{\;\;\;\;\gamma}_{(j)\delta} v_{(j)}^\delta \right]
       v_{(j)\gamma}
       - 2 \left( \mu_{(j)} + p_{(j)} \right) C_{\gamma\delta}
       v_{(j)}^\gamma v_{(j)}^\delta
       \right\}
       \Bigg\}.
\eea Thus, we have \bea
   & & \delta \mu
       = \sum_j \Big\{
       \delta \mu_{(j)}
       + \left( \mu_{(j)} + p_{(j)} \right) v_{(j)}^\alpha
       \left( v_{(j)\alpha} - v_\alpha \right)
   \nonumber \\
   & & \qquad
       + \left[ \left( \delta \mu_{(j)} + \delta p_{(j)} \right) v_{(j)}^\alpha
       + \Pi^{\;\;\;\;\alpha}_{(j)\beta} v_{(j)}^\beta
       - 2 \left( \mu_{(j)} + p_{(j)} \right)
       C^\alpha_\beta v_{(j)}^\beta
       \right]
       \left( v_{(j)\alpha} - v_\alpha \right)
       \Big\},
   \nonumber \\
   & & \delta p
       = \sum_j \Big\{
       \delta p_{(j)}
       + {1 \over 3} \left( \mu_{(j)} + p_{(j)} \right) v_{(j)}^\alpha
       \left( v_{(j)\alpha} - v_\alpha \right)
   \nonumber \\
   & & \qquad
       + {1 \over 3}\left[ \left( \delta \mu_{(j)} + \delta p_{(j)} \right)
       v_{(j)}^\alpha
       + \Pi^{\;\;\;\;\alpha}_{(j)\beta} v_{(j)}^\beta
       - 2 \left( \mu_{(j)} + p_{(j)} \right)
       C^\alpha_\beta v_{(j)}^\beta
       \right]
       \left( v_{(j)\alpha} - v_\alpha \right)
       \Big\},
   \nonumber \\
   & & \left( \mu + p \right) v_\alpha
       = \sum_j \Bigg\{ \left( \mu_{(j)} + p_{(j)} \right) v_{(j)\alpha}
       + \left( \delta \mu_{(j)} + \delta p_{(j)} \right)
       \left( v_{(j)\alpha} - v_\alpha \right)
       + \Pi^{\;\;\;\;\beta}_{(j)\alpha}
       \left( v_{(j)\beta} - v_\beta \right)
   \nonumber \\
   & & \qquad
       + \left( \mu_{(j)} + p_{(j)} \right) v_{(j)}^\beta v_{(j)\beta}
       \left( v_{(j)\alpha} - v_\alpha \right)
       - \left[
       {1 \over 2} \left( \mu_{(j)} + p_{(j)} \right)
       v_{(j)\alpha} \left( v_{(j)}^\beta + 3 v^\beta \right)
       + 2 \Pi_{(j)\alpha\beta} C^{\beta\gamma} \right]
       \left( v_{(j)\beta} - v_\beta \right)
       \Bigg\},
   \nonumber \\
   & & \Pi^\alpha_\beta
       = \sum_j \Bigg\{ \Pi^{\;\;\;\;\alpha}_{(j)\beta}
       + \left( \mu_{(j)} + p_{(j)} \right)  \left[ v_{(j)}^\alpha
       \left( v_{(j)\beta} - v_\beta \right)
       - {1 \over 3} \delta^\alpha_\beta v_{(j)}^\gamma
       \left( v_{(j)\gamma} - v_\gamma \right) \right]
   \nonumber \\
   & & \qquad
       + \left[ \left( \delta \mu_{(j)} + \delta p_{(j)} \right) v_{(j)}^\alpha
       - \Pi^{\;\;\;\;\alpha}_{(j)\gamma} v^\gamma \right]
       \left( v_{(j)\beta} - v_\beta \right)
       - {2 \over 3} \left( \mu_{(j)} + p_{(j)} \right) C^\alpha_\beta
       v_{(j)}^\gamma
       \left( v_{(j)\gamma} - v_\gamma \right)
   \nonumber \\
   & & \qquad
       - {1 \over 3} \delta^\alpha_\beta
       \left[ \left( \delta \mu_{(j)} + \delta p_{(j)} \right) v_{(j)}^\gamma
       + \Pi^{\;\;\;\;\gamma}_{(j)\delta} v_{(j)}^\delta
       - 2 \left( \mu_{(j)} + p_{(j)} \right) C^\gamma_\delta
       v_{(j)}^\delta \right]
       \left( v_{(j)\gamma} - v_\gamma \right)
       \Bigg\}.
   \label{total-fluid-quantities}
\eea

The ADM fluid quantities are introduced as \bea
   & & E \equiv \tilde n_a \tilde n_b \tilde T^{ab}, \quad
       J_\alpha \equiv - \tilde n_b \tilde T^b_\alpha, \quad
       S_{\alpha\beta} \equiv \tilde T_{\alpha\beta}, \quad
       S \equiv h^{\alpha\beta} S_{\alpha\beta}, \quad
       \bar S_{\alpha\beta} \equiv S_{\alpha\beta}
       - {1 \over 3} h_{\alpha\beta} S,
   \label{ADM-fluid-quantities}
\eea where tensor indices of $J_\alpha$ and $S_{\alpha\beta}$ are
based on $h_{\alpha\beta}$ as the metric. The four-vector $\tilde
n_a$ is a normal frame four-vector with $\tilde n_\alpha \equiv 0$.
Thus by setting $v_\alpha \equiv 0$ we have $\tilde u_a = \tilde
n_a$, and Eq.\ (\ref{u_a-v}) gives \bea
   & & \tilde n_\alpha \equiv 0,
   \nonumber \\
   & & \tilde n_0 = - a \left(
       1 + A - {1 \over 2} A^2
       + {1 \over 2} B^\alpha B_\alpha
       + {1 \over 2} A^3
       - {1 \over 2} A B^\alpha B_\alpha
       - C_{\alpha\beta} B^\alpha B^\beta \right),
   \nonumber \\
   & & \tilde n^\alpha = {1 \over a} \left(
       B^\alpha - A B^\alpha
       - 2 C^{\alpha\beta} B_\beta
       + {3 \over 2} A^2 B^\alpha
       + 2 A B^\beta C^\alpha_\beta
       - {1 \over 2} B^\alpha B^\beta B_\beta
       + 4 C^\alpha_\beta C^\beta_\gamma B^\gamma
       \right),
   \nonumber \\
   & & \tilde n^0 = {1 \over a} \left(
       1 - A + {3 \over 2} A^2
       - {1 \over 2} B^\alpha B_\alpha
       - {5 \over 2} A^3
       + {3 \over 2} A  B^\alpha B_\alpha
       + C_{\alpha\beta} B^\alpha B^\beta \right).
   \label{n_a}
\eea Using Eqs.\ (\ref{Tab-pert}),(\ref{n_a}), Eq.\
(\ref{ADM-fluid-quantities}) gives \bea
   & & E = \mu + \delta \mu
       + \left( \mu + p \right) v^\alpha v_\alpha
       + \left( \delta \mu + \delta p \right) v^\alpha v_\alpha
       - 2 \left( \mu + p \right) C_{\alpha\beta} v^\alpha v^\beta
       + \Pi_{\alpha\beta} v^\alpha v^\beta,
   \nonumber \\
   & & J_\alpha
       = a \left[ \left( \mu + p \right) v_\alpha
       + \left( \delta \mu + \delta p \right) v_\alpha
       + \Pi_{\alpha\beta} v^\beta
       + {1 \over 2} \left( \mu + p \right) v_\alpha v^\beta v_\beta
       - 2 \Pi_{\alpha\beta} C^\beta_\gamma v^\gamma \right],
   \nonumber \\
   & & S_{\alpha\beta}
       = a^2 \left[ p g^{(3)}_{\alpha\beta}
       + \delta p g^{(3)}_{\alpha\beta}
       + \Pi_{\alpha\beta}
       + 2 p C_{\alpha\beta}
       + \left( \mu + p \right) v_\alpha v_\beta
       + 2 \delta p C_{\alpha\beta}
       + \left( \delta \mu + \delta p \right) v_\alpha v_\beta
       \right],
   \nonumber \\
   & & S = 3 p + 3 \delta p
       + \left( \mu + p \right) v^\alpha v_\alpha
       + \left( \delta \mu + \delta p \right) v^\alpha v_\alpha
       + v^\alpha v^\beta \left[ \Pi_{\alpha\beta}
       - 2 \left( \mu + p \right) C_{\alpha\beta} \right],
   \nonumber \\
   & & \bar S_{\alpha\beta}
       = a^2 \Bigg\{
       \Pi_{\alpha\beta}
       + \left( \mu + p \right) \left( v_\alpha v_\beta
       - {1 \over 3} g^{(3)}_{\alpha\beta} v^\gamma v_\gamma
       \right)
       + \left( \delta \mu + \delta p \right) v_\alpha v_\beta
       - {2 \over 3} \left( \mu + p \right) C_{\alpha\beta} v^\gamma
       v_\gamma
   \nonumber \\
   & & \qquad
       - {1 \over 3} g^{(3)}_{\alpha\beta} \left[
       \left( \delta \mu
       + \delta p \right) v^\gamma v_\gamma
       + \Pi_{\gamma\delta} v^\gamma v^\delta
       - 2 \left( \mu + p \right) C_{\gamma\delta} v^\gamma v^\delta
       \right] \Bigg\}.
   \label{ADM-fluid-pert}
\eea The individual ADM fluid quantities can be found by replacing
\bea
   & & E, \quad
       J_\alpha, \quad
       S_{\alpha\beta}, \quad
       S, \quad
       \bar S_{\alpha\beta},
\eea with \bea
   & & E_{(i)}, \quad
       J_{(i)\alpha}, \quad
       S_{(i)\alpha\beta}, \quad
       S_{(i)}, \quad
       \bar S_{(i)\alpha\beta},
\eea and similarly for the fluid quantities and the energy-momentum
tensor as in Eqs.\ (\ref{total}),(\ref{individual}). From Eq.\
(\ref{Tab-i}) we have \bea
   & & E = \sum_j E_{(j)}, \quad
       J_\alpha = \sum_j J_{(j)\alpha}, \quad
       S_{\alpha\beta} = \sum_j S_{(j)\alpha\beta}, \quad
       S = \sum_j S_{(j)}, \quad
       \bar S_{\alpha\beta} = \sum_j \bar S_{(j)\alpha\beta}.
   \label{ADM-fluid-sum}
\eea

\subsection{Decomposition}

We decompose the metric into three perturbation types
\cite{Bardeen-1988} \bea
   & & A \equiv \alpha, \quad
       B_\alpha \equiv \beta_{,\alpha} + B^{(v)}_\alpha, \quad
       C_{\alpha\beta} \equiv \varphi g^{(3)}_{\alpha\beta}
       + \gamma_{,\alpha|\beta}
       + C^{(v)}_{(\alpha|\beta)}
       + C^{(t)}_{\alpha\beta},
\eea where superscripts $(v)$ and $(t)$ indicate the transverse
vector-type, and transverse-tracefree tensor-type perturbations,
respectively. Only to the linear-order perturbations in the
homogeneous-isotropic background, these three types of perturbations
decouple and evolve independently. We introduce \bea
   & & \chi \equiv a \left( \beta + c^{-1} a \dot \gamma \right), \quad
       \Psi^{(v)}_\alpha \equiv B^{(v)}_\alpha + c^{-1} a \dot
       C^{(v)}_\alpha,
\eea which are spatially gauge-invariant combinations to the linear
order \cite{Bardeen-1980}. We set \bea
   & & K \equiv - 3 H + \kappa.
\eea By using $\kappa$ we can avoid third-order expansion of the
trace of extrinsic curvature $K$. Identifying $\kappa$ with
Newtonian velocity variable later will be an important step in our
analysis, see Eqs.\ (\ref{identify-second}),(\ref{identify-third}).

For the fluid quantities we decompose \bea
   & & v_\alpha
       \equiv - v_{,\alpha} + v_{\alpha}^{(v)}, \quad
       \Pi_{\alpha\beta}
       \equiv {1 \over a^2} \left( \Pi_{,\alpha|\beta}
       - {1 \over 3} g^{(3)}_{\alpha\beta} \Delta \Pi \right)
       + {1 \over a} \Pi^{(v)}_{(\alpha|\beta)}
       + \Pi^{(t)}_{\alpha\beta},
   \nonumber \\
   & & v_{(i)\alpha}
       \equiv - v_{(i),\alpha} + v_{(i)\alpha}^{(v)}, \quad
       \Pi_{(i)\alpha\beta}
       \equiv {1 \over a^2} \left( \Pi_{(i),\alpha|\beta}
       - {1 \over 3} g^{(3)}_{\alpha\beta} \Delta \Pi_{(i)} \right)
       + {1 \over a} \Pi^{(v)}_{(i)(\alpha|\beta)}
       + \Pi^{(t)}_{(i)\alpha\beta}, \quad
       \delta I_{(i)\alpha} \equiv \delta I_{(i),\alpha}
       + \delta I_{(i)\alpha}^{(v)}.
\eea The perturbed fluid velocity variables $v$ and
$v_{\alpha}^{(v)}$ subtly differ from the ones introduced in
\cite{NL}, see \cite{second-order-multi}.

\subsection{Zero-pressure irrotational fluids in the comoving gauge}

The zero-pressure condition sets \bea
   & & p_{(i)} \equiv 0, \quad
       \delta p_{(i)} \equiv 0 \equiv \Pi_{(i)\alpha\beta}.
\eea The comoving gauge condition ($v \equiv 0$) and irrotational
condition ($v_\alpha^{(v)} = 0$) give \bea
   & & v_\alpha \equiv 0.
\eea Under these conditions, Eq.\ (\ref{total-fluid-quantities})
becomes \bea
   & & \delta \mu
       = \sum_j \left(
       \delta \mu_{(j)}
       + \mu_{(j)} v_{(j)}^\alpha v_{(j)\alpha}
       + \delta \mu_{(j)} v_{(j)}^\alpha v_{(j)\alpha}
       - 2 \mu_{(j)} C_{\alpha\beta} v_{(j)}^\alpha v_{(j)}^\beta \right),
   \nonumber \\
   & & \delta p
       = {1 \over 3} \sum_j \left(
       \mu_{(j)} v_{(j)}^\alpha v_{(j)\alpha}
       + \delta \mu_{(j)} v_{(j)}^\alpha v_{(j)\alpha}
       - 2 \mu_{(j)} C_{\alpha\beta} v_{(j)}^\alpha v_{(j)}^\beta
       \right),
   \nonumber \\
   & & 0
       = \sum_j \left( \mu_{(j)} v_{(j)\alpha}
       + \delta \mu_{(j)} v_{(j)\alpha}
       + {1 \over 2} \mu_{(j)} v_{(j)\alpha} v_{(j)}^\beta v_{(j)\beta}
       \right),
   \nonumber \\
   & & \Pi^\alpha_\beta
       = \sum_j \Bigg[ \mu_{(j)} \left( v_{(j)}^\alpha v_{(j)\beta}
       - {1 \over 3} \delta^\alpha_\beta v_{(j)}^\gamma v_{(j)\gamma} \right)
       + \delta \mu_{(j)} \left( v_{(j)}^\alpha v_{(j)\beta}
       - {1 \over 3} \delta^\alpha_\beta v_{(j)}^\gamma v_{(j)\gamma} \right)
   \nonumber \\
   & & \qquad
       - {2 \over 3} \mu_{(j)}
       \left( C^\alpha_\beta v_{(j)}^\gamma v_{(j)\gamma}
       - \delta^\alpha_\beta C^\gamma_\delta v_{(j)}^\delta v_{(j)\gamma} \right)
       \Bigg].
   \label{total-fluid-quantities-p=0}
\eea Notice that for the nonlinear perturbations, the collective
pressure $\delta p$ and anisotropic stress $\Pi_{\alpha\beta}$ no
longer vanish even for zero-pressure fluids; also, the collective
energy density $\delta \mu$ is no longer a simple sum of individual
component.

For the ADM fluid quantities, Eq.\ (\ref{ADM-fluid-pert}) becomes
\bea
   & & E = \mu + \delta \mu, \quad
       J_\alpha = 0, \quad
       S_{\alpha\beta} = a^2 \left( \delta p g^{(3)}_{\alpha\beta}
       + \Pi_{\alpha\beta} + 2 \delta p C_{\alpha\beta} \right),
       \quad
       S = 3 \delta p, \quad
       \bar S_{\alpha\beta} = a^2 \Pi_{\alpha\beta}.
   \label{ADM-fluid-p=0-pert}
\eea For the individual component, Eq.\ (\ref{ADM-fluid-pert}) gives
\bea
   & & E_{(i)} = \mu_{(i)} + \delta \mu_{(i)}
       + \mu_{(i)} v_{(i)}^\alpha v_{(i)\alpha}
       + \delta \mu_{(i)} v_{(i)}^\alpha v_{(i)\alpha}
       - 2 \mu_{(i)} C_{\alpha\beta} v_{(i)}^\alpha v_{(i)}^\beta,
   \nonumber \\
   & & J_{(i)\alpha}
       = a \left( \mu_{(i)} v_{(i)\alpha}
       + \delta \mu_{(i)} v_{(i)\alpha}
       + {1 \over 2} \mu_{(i)} v_{(i)\alpha} v_{(i)}^\beta
       v_{(i)\beta} \right),
   \nonumber \\
   & & S_{(i)\alpha\beta}
       = a^2 \left(
       \mu_{(i)} v_{(i)\alpha} v_{(i)\beta}
       + \delta \mu_{(i)} v_{(i)\alpha} v_{(i)\beta}
       \right),
   \nonumber \\
   & & S_{(i)} = \mu_{(i)} v_{(i)}^\alpha v_{(i)\alpha}
       + \delta \mu_{(i)} v_{(i)}^\alpha v_{(i)\alpha}
       - 2 \mu_{(i)} C_{\alpha\beta} v_{(i)}^\alpha v_{(i)}^\beta,
   \nonumber \\
   & & \bar S_{(i)\alpha\beta}
       = a^2 \Big[
       \mu_{(i)} \left( v_{(i)\alpha} v_{(i)\beta}
       - {1 \over 3} g^{(3)}_{\alpha\beta} v_{(i)}^\gamma
       v_{(i)\gamma}
       \right)
       + \delta \mu_{(i)} v_{(i)\alpha} v_{(i)\beta}
       - {2 \over 3} \mu_{(i)} C_{\alpha\beta} v_{(i)}^\gamma
       v_{(i)\gamma}
   \nonumber \\
   & & \qquad
       - {1 \over 3} g^{(3)}_{\alpha\beta} \left(
       \delta \mu_{(i)}  v_{(i)}^\gamma v_{(i)\gamma}
       - 2 \mu_{(i)} C_{\gamma\delta} v_{(i)}^\gamma v_{(i)}^\delta
       \right) \Big].
   \label{ADM-fluid-p=0-pert-i}
\eea

%
%
\section{Second-order equations: summary}
                                        \label{sec:second-order}

In this section we summarize our previous work of second-order
perturbations in zero-pressure, irrotational, but multi-component
fluids \cite{second-order-multi}. The results show that, to the
second order, effectively the relativistic equations coincide with
the Newtonian ones even in the multi-component situation. This
provides a reason to go to the third order in relativistic
perturbation in order to find out pure general relativistic
deviations from Newton's theory. Since the zero-pressure Newtonian
perturbation equations have only quadratic order nonlinearity, any
non-vanishing third-order terms in relativistic analysis can be
regarded as a pure general relativistic correction terms. In
\cite{third-order} we presented such third-order correction terms in
a single component case. In the next section we will derive the
third-order correction terms appearing in the multi-component
situation. As the third-order analysis closely follows the case of
second-order, in the following we will derive the second-order
equations directly from the ADM equations presented in Sec.\
\ref{sec:covariant-eqs}.

We consider zero-pressure multi-fluids, thus \bea
   & & p_{(i)} \equiv 0, \quad
       \delta p_{(i)} \equiv 0 \equiv \Pi_{(i)\alpha\beta}.
\eea The irrotational condition and the temporal comoving gauge
condition lead to $v_\alpha = 0$, thus we have $J_\alpha = 0$; as
the spatial gauge condition we take $\gamma \equiv 0$, thus $\beta =
\chi/a$. To the second-order perturbations, Eq.\
(\ref{Mom-conservation-ADM}) gives \bea
   & & \alpha = - {1 \over 2 a^2} \chi^{,\alpha} \chi_{,\alpha}
       - \sum_j {\mu_{(j)} \over \mu} \left[
       {1 \over 2} v_{(j)}^\alpha v_{(j)\alpha}
       + \Delta^{-1} \nabla_\alpha \left(
       v_{(j)}^\alpha v^\beta_{(j)|\beta} \right) \right],
\eea thus \bea
   & & N = a
       - a \sum_j {\mu_{(j)} \over \mu} \left[
       {1 \over 2} v_{(j)}^\alpha v_{(j)\alpha}
       + \Delta^{-1} \nabla_\alpha \left(
       v_{(j)}^\alpha v^\beta_{(j)|\beta} \right) \right].
\eea Using Eqs.\
(\ref{ADM-fluid-p=0-pert}),(\ref{ADM-fluid-p=0-pert-i}), Eqs.\
(\ref{E-conservation-ADM}),(\ref{Trace-prop-ADM}),(\ref{E-conservation-ADM-i}),(\ref{Mom-conservation-ADM-i})
give \bea
   & & \dot \delta
       - \kappa
       =
       - {c \over a^2} \delta_{,\alpha} \chi^{,\alpha}
       + \delta \kappa
       + {1 \over 2} H \sum_j
       \mu_{(j)} v_{(j)}^\alpha v_{(j)\alpha}
       + 3 H \sum_j {\mu_{(j)} \over \mu} \Delta^{-1} \nabla_\alpha \left(
       v_{(j)}^\alpha v^\beta_{(j)|\beta} \right),
   \label{E-conservation-zero-pressure-irrotation-total-CG} \\
   & & \dot \kappa + 2 H \kappa
       - 4 \pi G \varrho \delta
       =
       - {c \over a^2} \kappa_{,\alpha} \chi^{,\alpha}
       + {1 \over 3} \kappa^2
       + \left( \dot C^{(t)\alpha\beta}
       + {c \over a^2} \chi^{,\alpha|\beta} \right)
       \left( \dot C^{(t)}_{\alpha\beta}
       + {c \over a^2} \chi_{,\alpha|\beta} \right)
       - {1 \over 3} \left( c {\Delta \over a^2} \chi \right)^2
   \nonumber \\
   & & \qquad
       + {1 \over 2} \left( 3 \dot H + 8 \pi G \varrho + c^2 {\Delta \over a^2} \right)
       \sum_j \mu_{(j)} v_{(j)}^\alpha v_{(j)\alpha}
       + \left( 3 \dot H + c^2 {\Delta \over a^2} \right)
       \sum_j {\mu_{(j)} \over \mu} \Delta^{-1} \nabla_\alpha \left(
       v_{(j)}^\alpha v^\beta_{(j)|\beta} \right),
   \label{Raychaudhury-eq-zero-pressure-irrotation-CG} \\
   & & \dot \delta_{(i)}
       - \kappa
       + c {1 \over a} \left[ \left( 1 + \delta_{(i)} \right)
       v_{(i)}^{\alpha} \right]_{|\alpha}
       =
       - {c \over a^2} \delta_{(i),\alpha} \chi^{,\alpha}
       + \delta_{(i)} \kappa
       + H v_{(i)\alpha} v_{(i)}^{\alpha}
   \nonumber \\
   & & \qquad
       - {c \over a} \left( \varphi^{,\alpha} v_{(i)\alpha}
       - 2 \varphi v^\alpha_{(i)|\alpha}
       - 2 v_{(i)}^{\alpha|\beta} C^{(t)}_{\alpha\beta} \right)
       + {3 \over 2} H \sum_j \mu_{(j)} v_{(j)}^\alpha v_{(j)\alpha}
       + 3 H \sum_j {\mu_{(j)} \over \mu} \Delta^{-1} \nabla_\alpha \left(
       v_{(j)}^\alpha v^\beta_{(j)|\beta} \right),
   \label{E-conservation-zero-pressure-irrotation-CG} \\
   & & {1 \over a} \left[ a \left( 1 + \delta_{(i)} \right)
       v_{(i)\alpha} \right]^\cdot
       =
       - {c \over a^2} \left( v_{(i)\beta} \chi^{,\beta} \right)_{,\alpha}
       + \kappa v_{(i)\alpha}
       - {c \over a} \left( v_{(i)\alpha} v_{(i)}^{\beta} \right)_{|\beta}
   \nonumber \\
   & & \qquad
       + {c \over 2 a} \sum_j \mu_{(j)}
       \nabla_\alpha \left( v_{(j)}^\beta v_{(j)\beta} \right)
       + {c \over a} \sum_j {\mu_{(j)} \over \mu} \nabla_\alpha
       \Delta^{-1} \nabla_\gamma \left(
       v_{(j)}^\gamma v^\beta_{(j)|\beta} \right).
   \label{Mom-conservation-zero-pressure-irrotation-CG}
\eea From Eqs.
(\ref{E-conservation-zero-pressure-irrotation-total-CG}),(\ref{Raychaudhury-eq-zero-pressure-irrotation-CG}),
and Eqs.\
(\ref{Raychaudhury-eq-zero-pressure-irrotation-CG})-(\ref{Mom-conservation-zero-pressure-irrotation-CG})
we can derive \bea
   & & {1 \over a^2} \left[ a^2 \left( \dot \delta
       + {c \over a^2} \delta_{,\alpha} \chi^{,\alpha} \right)
       \right]^\cdot
       - 4 \pi G \varrho \delta \left( 1 + \delta \right)
       = - {c \over a^2} \kappa_{,\alpha} \chi^{,\alpha}
       + {4 \over 3} \kappa^2
   \nonumber \\
   & & \qquad
       + \left( \dot C^{(t)\alpha\beta}
       + {c \over a^2} \chi^{,\alpha|\beta} \right)
       \left( \dot C^{(t)}_{\alpha\beta}
       + {c \over a^2} \chi_{,\alpha|\beta} \right)
       - {1 \over 3} \left( c {\Delta \over a^2} \chi \right)^2
   \nonumber \\
   & & \qquad
       + \left( 2 \dot H + 4 \pi G \varrho
       + {1 \over 2} c^2 {\Delta \over a^2}\right)
       \sum_j \mu_{(j)} v_{(j)}^\alpha v_{(j)\alpha}
       + \left( 6 \dot H + c^2 {\Delta \over a^2} \right)
       \sum_j {\mu_{(j)} \over \mu} \Delta^{-1} \nabla_\alpha \left(
       v_{(j)}^\alpha v^\beta_{(j)|\beta} \right),
   \label{ddot-eq-CG} \\
   & & {1 \over a^2} \left[ a^2 \left( \dot \delta_{(i)}
       + {c \over a^2} \delta_{(i),\alpha} \chi^{,\alpha} \right)
       \right]^\cdot
       - 4 \pi G \varrho \delta \left( 1 + \delta_{(i)} \right)
       = - {c \over a^2} \kappa_{,\alpha} \chi^{,\alpha}
       + {4 \over 3} \kappa^2
   \nonumber \\
   & & \qquad
       + \left( \dot C^{(t)\alpha\beta}
       + {c \over a^2} \chi^{,\alpha|\beta} \right)
       \left( \dot C^{(t)}_{\alpha\beta}
       + {c \over a^2} \chi_{,\alpha|\beta} \right)
       - {1 \over 3} \left( c {\Delta \over a^2} \chi \right)^2
       - {c \over a} \left[ \left( \kappa + \dot \varphi \right)^{,\alpha}
       v_{(i)\alpha}
       + 2 \left( \kappa - \dot \varphi \right) v^\alpha_{(i)|\alpha}
       \right]
   \nonumber \\
   & & \qquad
       + c^2 {\Delta \over a^3} \left( v_{(i)\alpha} \chi^{,\alpha} \right)
       + {c^2 \over a^2} \left( v_{(i)}^\alpha v_{(i)}^\beta \right)_{|\alpha\beta}
       + \dot H v_{(i)\alpha} v_{(i)}^{\alpha}
       + {2 c \over a} v_{(i)}^{\alpha|\beta} \dot C^{(t)}_{\alpha\beta}
   \nonumber \\
   & & \qquad
       + \left( 3 \dot H + 4 \pi G \varrho \right)
       \sum_j \mu_{(j)} v_{(j)}^\alpha v_{(j)\alpha}
       + 6 \dot H \sum_j {\mu_{(j)} \over \mu} \Delta^{-1} \nabla_\alpha \left(
       v_{(j)}^\alpha v^\beta_{(j)|\beta} \right).
   \label{ddot-eq-CG-i}
\eea We have recovered $c$ using \bea
   & & [G \varrho] = T^{-2}, \quad
       [p] = [\mu] = [\varrho c^2], \quad
       [a] = L, \quad
       [R^{(3)}] = 1, \quad
       [\nabla] = 1,
   \nonumber \\
   & &
       [\varphi]
       = [\beta]
       = [\gamma]
       = [C^{(t)}_{\alpha\beta}]
       = [v_\alpha] = [v_{(i)\alpha}]
       = [\delta] = [\delta_{(i)}] = 1, \quad
       [\chi] = L, \quad
       [\kappa] = T^{-1},
\eea where $R^{(3)} = 6 \bar K$ with normalized $\bar K = 0$ or $\pm
1$.

Without the rotational mode, we introduce \bea
   & & v_{(i)\alpha} \equiv - v_{(i),\alpha}.
\eea Since we used the comoving gauge condition, to the linear
order, we have \bea
   & & v_{(i)}
       = v_{(i)v}
       \equiv v_{(i)} - v
       = v_{(i)\chi} - v_\chi,
\eea where $v_\chi \equiv v - \chi/a$ and $v_{(i)\chi} \equiv
v_{(i)} - \chi/a$ to the linear order. To the linear order,
Newtonian velocity perturbation variables are introduced as
\cite{HN-multi-CQG} \bea
   & & {\bf u} \equiv \nabla u
       \equiv - c \nabla v_\chi, \quad
       {\bf u}_i \equiv \nabla u_i
       \equiv - c \nabla v_{(i)\chi},
\eea thus \bea
   & & {\bf v}_{(i)} \equiv - \nabla v_{(i)}
       = {1 \over c} \left( {\bf u}_i - {\bf u}
       \right).
   \label{v_i}
\eea Dimensions are \bea
   & & [v] = [v_{(i)}] = 1, \quad
       [{\bf u}] = [{\bf u}_i] = [u] = [u_i] = [c] = L/T.
\eea Thus, we have \bea
   & & {\delta p \over \mu}
       = {1 \over 3} \sum_j {\varrho_j \over \varrho}
       {1 \over c^2} | {\bf u}_j - {\bf u} |^2,
   \nonumber \\
   & & \sum_j {\mu_{(j)} \over \mu}
       \Delta^{-1} \nabla_\alpha \left(
       v_{(j)}^\alpha v^\beta_{(j)|\beta} \right)
       = \sum_j {\varrho_j \over \varrho} {1 \over c^2} \Delta^{-1}
       \nabla \cdot \left[ \left( {\bf u}_j - {\bf u} \right)
       \nabla \cdot \left( {\bf u}_j - {\bf u} \right) \right].
\eea

\subsection{Newtonian correspondence}

We {\it assume} a flat background, thus $R^{(3)} = 0$. Thus, we have
$\dot \varphi_v = 0$ to the linear order even in the presence of the
cosmological constant and multiple components
\cite{Hwang-Noh-Newtonian,HN-multi-CQG}, see Eq.\
(\ref{dot-varphi-eq}). To the second order, we {\it identify} the
Newtonian perturbation variables $\delta$, $\delta_i$, ${\bf u}$,
and ${\bf u}_i$ as \bea
   & & \kappa_v \equiv - {1 \over a} \nabla \cdot {\bf u}, \quad
       {\bf u} \equiv \nabla u, \quad
       {\bf v}_{(i)v} \equiv {1 \over c} \left( {\bf u}_i - {\bf u}
       \right), \quad
       \delta \equiv \delta_v, \quad
       \delta_i \equiv \delta_{(i)v}.
   \label{identify-second}
\eea Comparison of the consequent relativistic equations with the
Newtonian equations will show apparently why these identifications
are the proper ones. Examination of Eqs.\
(\ref{E-conservation-zero-pressure-irrotation-total-CG})-(\ref{ddot-eq-CG-i})
shows that we need $\chi (= \chi_v)$ only to the linear order. From
the momentum constraint equation in Eq.\ (\ref{Mom-constraint-ADM}),
see Eq. (197) of \cite{NL}, we have \bea
   & & \kappa_v = - c {\Delta \over a^2} \chi_v,
\eea thus \bea
   & & \chi_v \equiv {a \over c} u,
\eea to the linear order.

Equations
(\ref{E-conservation-zero-pressure-irrotation-total-CG})-(\ref{ddot-eq-CG-i})
become \bea
   & & \dot \delta + {1 \over a} \nabla \cdot {\bf u}
       = - {1 \over a} \nabla \cdot \left( \delta {\bf u} \right)
       + H \sum_j {\varrho_j \over \varrho} {1 \over c^2}
       \left\{
       {1 \over 2} |{\bf u}_j - {\bf u}|^2
       + 3 \Delta^{-1} \nabla \cdot \left[
       \left( {\bf u}_j - {\bf u} \right)
       \nabla \cdot \left( {\bf u}_j - {\bf u} \right) \right]
       \right\},
   \label{dot-delta-second} \\
   & & {1 \over a} \nabla \cdot \left( \dot {\bf u} + H {\bf u}
       \right)
       + 4 \pi G \varrho \delta
       = - {1 \over a^2} \nabla \cdot \left( {\bf u} \cdot \nabla
       {\bf u} \right)
       - \dot C^{(t)\alpha\beta} \left(
       {2 \over a} u_{\alpha|\beta} + \dot C^{(t)}_{\alpha\beta}
       \right)
   \nonumber \\
   & & \qquad
       + \sum_j {\varrho_j \over \varrho} {1 \over c^2}
       \left\{ {1 \over 2} \left( 4 \pi G \varrho - c^2 {\Delta
       \over a^2} \right)
       |{\bf u}_j - {\bf u}|^2
       + \left( 12 \pi G \varrho - c^2 {\Delta
       \over a^2} \right) \Delta^{-1} \nabla \cdot \left[
       \left( {\bf u}_j - {\bf u} \right)
       \nabla \cdot \left( {\bf u}_j - {\bf u} \right) \right]
       \right\},
   \label{dot-u-second} \\
   & & \dot \delta_i
       + {1 \over a} \nabla \cdot {\bf u}_i
       = - {1 \over a} \nabla \cdot \left( \delta_i {\bf u}_i
       \right)
       + {1 \over a} \left[ 2 \varphi \nabla \cdot
       \left( {\bf u}_i - {\bf u} \right)
       - \left( {\bf u}_i - {\bf u} \right) \cdot \nabla \varphi
       + 2 \left( u^\alpha_i - u^\alpha \right)^{|\beta}
       C^{(t)}_{\alpha\beta} \right]
   \nonumber \\
   & & \qquad
       + H {1 \over c^2} |{\bf u}_i - {\bf u}|^2
       + 3 H \sum_j {\varrho_j \over \varrho} {1 \over c^2}
       \left\{
       {1 \over 2} |{\bf u}_j - {\bf u}|^2
       + \Delta^{-1} \nabla \cdot \left[
       \left( {\bf u}_j - {\bf u} \right)
       \nabla \cdot \left( {\bf u}_j - {\bf u} \right) \right]
       \right\},
   \label{dot-delta-i-second} \\
   & & {1 \over a} \nabla \cdot \left( \dot {\bf u}_i
       + H {\bf u}_i \right)
       + 4 \pi G \varrho \delta
       = - {1 \over a^2} \nabla \cdot \left( {\bf u}_i \cdot
       \nabla {\bf u}_i \right)
       - \dot C^{(t)\alpha\beta} \left(
       {2 \over a} u_{\alpha|\beta} + \dot C^{(t)}_{\alpha\beta}
       \right)
   \nonumber \\
   & & \qquad
       + 4 \pi G \sum_j \varrho_j {1 \over c^2}
       \left\{
       {1 \over 2} |{\bf u}_j - {\bf u}|^2
       + 3 \Delta^{-1} \nabla \cdot \left[
       \left( {\bf u}_j - {\bf u} \right)
       \nabla \cdot \left( {\bf u}_j - {\bf u} \right) \right]
       \right\},
   \label{dot-u-i-second} \\
   & & {1 \over a^2} \left( a^2 \dot \delta \right)^\cdot
       - 4 \pi G \varrho \delta
       = - {1 \over a^2} \left[ a \nabla \cdot \left( \delta {\bf u}
       \right) \right]^\cdot
       + {1 \over a^2} \nabla \cdot \left( {\bf u} \cdot \nabla
       {\bf u} \right)
       + \dot C^{(t)\alpha\beta} \left(
       {2 \over a} u_{\alpha|\beta} + \dot C^{(t)}_{\alpha\beta}
       \right)
   \nonumber \\
   & & \qquad
       - \sum_j {\varrho_j \over \varrho} {1 \over c^2}
       \left\{ \left( 4 \pi G \varrho
       - {c^2 \over 2} {\Delta \over a^2} \right) |{\bf u}_j - {\bf u}|^2
       + \left( 24 \pi G \varrho
       - c^2 {\Delta \over a^2} \right)
       \Delta^{-1} \nabla \cdot \left[
       \left( {\bf u}_j - {\bf u} \right)
       \nabla \cdot \left( {\bf u}_j - {\bf u} \right) \right]
       \right\},
   \label{ddot-delta-second}\\
   & & {1 \over a^2} \left( a^2 \dot \delta_i \right)^\cdot
       - 4 \pi G \varrho \delta
       = - {1 \over a^2} \left[ a \nabla \cdot \left( \delta_i {\bf
       u}_i \right) \right]^\cdot
       + {1 \over a^2} \nabla \cdot \left( {\bf u}_i \cdot \nabla
       {\bf u}_i \right)
       + \dot C^{(t)}_{\alpha\beta} \left(
       {2 \over a} u_i^{\alpha|\beta} + \dot C^{(t)\alpha\beta}
       \right)
   \nonumber \\
   & & \qquad
       + {1 \over a^2} \left\{ \Delta \left[ {\bf u} \cdot \left(
       {\bf u}_i - {\bf u} \right) \right]
       - \nabla \cdot \left[ \left( {\bf u}_i - {\bf u} \right)
       \cdot \nabla {\bf u}
       + {\bf u} \cdot \nabla \left( {\bf u}_i - {\bf u} \right)
       \right] \right\}
   \nonumber \\
   & & \qquad
       - {4 \pi G \varrho \over c^2} |{\bf u}_i - {\bf u}|^2
       - 8 \pi G \sum_j \varrho_j {1 \over c^2} \left\{
       |{\bf u}_j - {\bf u}|^2
       + 3 \Delta^{-1} \nabla \cdot \left[
       \left( {\bf u}_j - {\bf u} \right)
       \nabla \cdot \left( {\bf u}_j - {\bf u} \right) \right]
       \right\}.
   \label{ddot-delta-i-second}
\eea

In \cite{second-order-multi} we have shown that, to the linear
order, $({\bf u}_i - {\bf u})$ simply decays \bea
   & & {\bf u}_i - {\bf u} \propto {1 \over a},
   \label{u_i-u}
\eea in an expanding phase. Thus, ignoring quadratic combination of
$({\bf u}_i - {\bf u})$ terms, we have \bea
   & & \dot \delta + {1 \over a} \nabla \cdot {\bf u}
       = - {1 \over a} \nabla \cdot \left( \delta {\bf u} \right),
   \label{dot-delta-second-2} \\
   & & {1 \over a} \nabla \cdot \left( \dot {\bf u} + H {\bf u}
       \right)
       + 4 \pi G \varrho \delta
       = - {1 \over a^2} \nabla \cdot \left( {\bf u} \cdot \nabla
       {\bf u} \right)
       - \dot C^{(t)\alpha\beta} \left(
       {2 \over a} u_{\alpha|\beta} + \dot C^{(t)}_{\alpha\beta}
       \right),
   \label{dot-u-second-2} \\
   & & {1 \over a^2} \left( a^2 \dot \delta \right)^\cdot
       - 4 \pi G \varrho \delta
       = - {1 \over a^2} \left[ a \nabla \cdot \left( \delta {\bf u}
       \right) \right]^\cdot
       + {1 \over a^2} \nabla \cdot \left( {\bf u} \cdot \nabla
       {\bf u} \right)
       + \dot C^{(t)\alpha\beta} \left(
       {2 \over a} u_{\alpha|\beta} + \dot C^{(t)}_{\alpha\beta}
       \right),
   \label{ddot-delta-second-2}
\eea and \bea
   & & \dot \delta_i
       + {1 \over a} \nabla \cdot {\bf u}_i
       = - {1 \over a} \nabla \cdot \left( \delta_i {\bf u}_i
       \right)
       + {1 \over a} \left[ 2 \varphi \nabla \cdot
       \left( {\bf u}_i - {\bf u} \right)
       - \left( {\bf u}_i - {\bf u} \right) \cdot \nabla \varphi
       + 2 \left( u^\alpha_i - u^\alpha \right)^{|\beta}
       C^{(t)}_{\alpha\beta} \right],
   \label{dot-delta-i-second-2} \\
   & & {1 \over a} \nabla \cdot \left( \dot {\bf u}_i
       + H {\bf u}_i \right)
       + 4 \pi G \varrho \delta
       = - {1 \over a^2} \nabla \cdot \left( {\bf u}_i \cdot
       \nabla {\bf u}_i \right)
       - \dot C^{(t)\alpha\beta} \left(
       {2 \over a} u_{\alpha|\beta} + \dot C^{(t)}_{\alpha\beta}
       \right),
   \label{dot-u-i-second-2} \\
   & & {1 \over a^2} \left( a^2 \dot \delta_i \right)^\cdot
       - 4 \pi G \varrho \delta
       = - {1 \over a^2} \left[ a \nabla \cdot \left( \delta_i {\bf
       u}_i \right) \right]^\cdot
       + {1 \over a^2} \nabla \cdot \left( {\bf u}_i \cdot \nabla
       {\bf u}_i \right)
       + \dot C^{(t)}_{\alpha\beta} \left(
       {2 \over a} u_i^{\alpha|\beta} + \dot C^{(t)\alpha\beta}
       \right)
   \nonumber \\
   & & \qquad
       + {1 \over a^2} \left\{ \Delta \left[ {\bf u} \cdot \left(
       {\bf u}_i - {\bf u} \right) \right]
       - \nabla \cdot \left[ \left( {\bf u}_i - {\bf u} \right)
       \cdot \nabla {\bf u}
       + {\bf u} \cdot \nabla \left( {\bf u}_i - {\bf u} \right)
       \right] \right\}.
   \label{ddot-delta-i-second-2}
\eea Equations
(\ref{dot-delta-second-2})-(\ref{ddot-delta-second-2}) coincide with
the density and velocity perturbation equations of a single
component medium \cite{second-order}; thus, except for the
contribution from gravitational waves, these equations coincide with
ones in the Newtonian context. In the Newtonian context, Eqs.\
(\ref{dot-delta-second-2})-(\ref{ddot-delta-second-2}) without the
gravitational waves were presented in \cite{Peebles-1980} which are
valid to fully nonlinear order. To the linear order, Eq.\
(\ref{ddot-delta-second-2}) was derived by Lifshitz
\cite{Lifshitz-1946} in the synchronous gauge, and by Nariai
\cite{Nariai-1969} in the comoving gauge. In the zero-pressure
medium the free-falling object is also comoving, thus we can impose
both the synchronous gauge and the comoving gauge simultaneously. In
\cite{second-order-SG} we compared subtle differences of the
second-order perturbation equations in the synchronous gauge with
the ones in the comoving gauge.

If we further ignore $({\bf u}_i - {\bf u})$ terms appearing in the
pure second-order combinations, Eqs.\
(\ref{dot-delta-i-second-2})-(\ref{ddot-delta-i-second-2}) become
\bea
   & & \dot \delta_i
       + {1 \over a} \nabla \cdot {\bf u}_i
       = - {1 \over a} \nabla \cdot \left( \delta_i {\bf u}_i
       \right),
   \label{dot-delta-i-second-3} \\
   & & {1 \over a} \nabla \cdot \left( \dot {\bf u}_i
       + H {\bf u}_i \right)
       + 4 \pi G \varrho \delta
       = - {1 \over a^2} \nabla \cdot \left( {\bf u}_i \cdot
       \nabla {\bf u}_i \right)
       - \dot C^{(t)\alpha\beta} \left(
       {2 \over a} u_{\alpha|\beta} + \dot C^{(t)}_{\alpha\beta}
       \right),
   \label{dot-u-i-second-3} \\
   & & {1 \over a^2} \left( a^2 \dot \delta_i \right)^\cdot
       - 4 \pi G \varrho \delta
       = - {1 \over a^2} \left[ a \nabla \cdot \left( \delta_i {\bf
       u}_i \right) \right]^\cdot
       + {1 \over a^2} \nabla \cdot \left( {\bf u}_i \cdot \nabla
       {\bf u}_i \right)
       + \dot C^{(t)}_{\alpha\beta} \left(
       {2 \over a} u_i^{\alpha|\beta} + \dot C^{(t)\alpha\beta}
       \right),
   \label{ddot-delta-i-second-3}
\eea which coincide with the Newtonian equations except for the
contributions from the gravitational waves. In this context, except
for the contribution from gravitational waves, the above equations
coincide with ones in the Newtonian context even in the
multi-component case. Therefore, we have shown the
relativistic/Newtonian correspondence, except for the contributions
from the gravitational waves, to the second-order perturbations in
the case of multi-component, zero-pressure, irrotational fluids
assuming a flat background.

%
%
\section{Third-order equations}
                                       \label{sec:third-order-equations}

We {\it consider} irrotational fluids, thus ignore all vector-type
perturbations. As all three-types of perturbations are generally
coupled in nonlinear perturbations, apparently this is an important
assumption we make in this section. In an expanding phase, however,
the linear order rotational perturbations have only decaying mode
due to the angular-momentum conservation. Effects of rotational
perturbations to the second-order perturbations are considered in
our accompanying work in \cite{second-order-multi}.

We {\it take} the temporal comoving gauge \bea
   & & v \equiv 0.
\eea Together with the irrotational condition $v_\alpha^{(v)} \equiv
0$ we have $v_\alpha = 0$. Equation (\ref{ADM-fluid-pert}) shows
that this leads to $J_\alpha = 0$ for general fluids. As the spatial
gauge condition we {\it take} \bea
   & & \gamma \equiv 0.
\eea In \cite{NL} we have shown that these gauge conditions fix the
space-time gauge transformation properties completely to all orders
in perturbations. Thus, each perturbation variable under these gauge
conditions has a corresponding unique gauge-invariant combination,
and can be equivalently regarded as {\it gauge-invariant} one to all
orders in perturbations.

We consider zero-pressure fluids, thus \bea
   & & p_{(i)} = 0, \quad
       \delta p_{(i)} \equiv 0 \equiv \Pi_{(i)\alpha\beta}.
\eea The collective fluid quantities are nontrivial and are
presented in Eq.\ (\ref{total-fluid-quantities-p=0}).

In our previous study of second-order perturbations, summarized in a
previous section, we showed that ${\cal O} (|v_{(i)\alpha}|^2)$
terms simply correspond to pure decaying mode in expanding phase. We
showed that by ignoring these terms we recover complete
relativistic/Newtonian correspondence even in the multi-component
case. Based on this observation, in our third order calculation in
the following we will {\it ignore} ${\cal O} (|v_{(i)\alpha}|^2)$
terms. If we {\it ignore} ${\cal O} (|v_{(i)\alpha}|^2)$ terms, the
collective fluid quantities in Eq.\
(\ref{total-fluid-quantities-p=0}) give \bea
   & & \delta \mu
       = \sum_j \delta \mu_{(j)}, \quad
       \delta p = 0, \quad
       0 = \sum_j \left( \mu_{(j)} v_{(j)\alpha}
       + \delta \mu_{(j)} v_{(j)\alpha} \right), \quad
       \Pi^\alpha_\beta = 0,
\eea and the ADM fluid quantities in Eqs.\
(\ref{ADM-fluid-p=0-pert}),(\ref{ADM-fluid-p=0-pert-i}) become \bea
   & & E = \mu + \delta \mu, \quad
       J_\alpha = 0, \quad
       S_{\alpha\beta} = 0, \quad
       S = 0, \quad
       \bar S_{\alpha\beta} = 0,
   \nonumber \\
   & & E_{(i)} = \mu_{(i)} + \delta \mu_{(i)}, \quad
       J_{(i)\alpha}
       = a \left( \mu_{(i)} v_{(i)\alpha}
       + \delta \mu_{(i)} v_{(i)\alpha} \right), \quad
       S_{(i)\alpha\beta} = 0, \quad
       S_{(i)} = 0, \quad
       \bar S_{(i)\alpha\beta} = 0.
   \label{ADM-fluid-third}
\eea Equation (\ref{Mom-conservation-ADM}) gives \bea
   & & N = a (t),
\eea thus, \bea
   & & \alpha = - {1 \over 2 a^2} \chi^{,\alpha} \chi_{,\alpha}
       \left( 1 - 2 \varphi \right)
       + {1 \over a^2} \chi^{,\alpha} \chi^{,\beta}
       C^{(t)}_{\alpha\beta}.
   \label{alpha-third}
\eea Using Eq.\ (\ref{ADM-fluid-third}), Eqs.\
(\ref{E-conservation-ADM}),(\ref{Trace-prop-ADM}),(\ref{E-conservation-ADM-i}),(\ref{Mom-conservation-ADM-i})
give \bea
   & & \left( {\dot \mu \over \mu} + 3 H \right) \left( 1 + \delta \right)
       + \dot \delta
       - \kappa
       = - {c \over a^2} \delta_{,\alpha} \chi^{,\alpha}
       \left( 1 - 2 \varphi \right)
       + \kappa \delta
       + {2 c \over a^2} \delta^{,\alpha} \chi^{,\beta}
       C^{(t)}_{\alpha\beta},
   \label{delta-eq-third} \\
   & & - \left[ 3 \dot H + 3 H^2 + 4 \pi G \varrho - \Lambda c^2 \right]
       + \dot \kappa + 2 H \kappa - 4 \pi G \varrho \delta
       = - {c \over a^2} \kappa_{,\alpha} \chi^{,\alpha}
       \left( 1 - 2 \varphi \right)
       + {1 \over 3} \kappa^2
       + {2 c \over a^2} \kappa^{,\alpha}
       \chi^{,\beta} C^{(t)}_{\alpha\beta}
   \nonumber \\
   & & \qquad
       + \left( {c \over a^2} \chi^{,\alpha|\beta}
       - {c \over 3} g^{(3)\alpha\beta} {\Delta \over a^2} \chi
       + \dot C^{(t)\alpha\beta} \right)
       \Bigg[
       \left( {c \over a^2} \chi_{,\alpha|\beta}
       + \dot C^{(t)}_{\alpha\beta} \right)
       \left( 1 - 4 \varphi \right)
   \nonumber \\
   & & \qquad
       - {4 c \over a^2} \chi_{,\alpha} \varphi_{,\beta}
       - {2 c \over a^2} \chi^{,\gamma} \left( 2
       C^{(t)}_{\gamma\alpha|\beta}
       - C^{(t)}_{\alpha\beta|\gamma} \right)
       - 4 C^{(t)}_{\beta\gamma}
       \left( {c \over a^2} \chi^{,\gamma}_{\;\;\;|\alpha}
       + \dot C^{(t)\gamma}_{\;\;\;\;\;\alpha} \right)
       - 4 \dot \varphi C^{(t)}_{\alpha\beta} \Bigg],
   \label{kappa-eq-third} \\
   & & \left( {\dot \mu_{(i)} \over \mu_{(i)}} + 3 H \right)
       \left( 1 + \delta_{(i)} \right)
       + \dot \delta_{(i)} - \kappa
       + {c \over a} \left[ \left( 1 + \delta_{(i)} \right)
       v_{(i)}^\alpha \right]_{|\alpha}
       \left( 1 - 2 \varphi + 4 \varphi^2 \right)
   \nonumber \\
   & & \qquad
       = - {c \over a^2} \delta_{(i),\alpha} \chi^{,\alpha}
       \left( 1 - 2 \varphi \right)
       + \kappa \delta_{(i)}
       - {c \over a} \varphi_{,\alpha} v_{(i)}^{\alpha}
       \left( 1 - 4 \varphi \right)
       + {2 c \over a} C^{(t)}_{\alpha\beta} v_{(i)}^{\alpha|\beta}
       \left( 1 - 4 \varphi \right)
       + {2 c \over a^2} \delta_{(i)}^{,\alpha} \chi^{,\beta}
       C^{(t)}_{\alpha\beta}
   \nonumber \\
   & & \qquad
       + {c \over a} \left[ - \varphi^{,\alpha}
       \left( \delta_{(i)} v_{(i)\alpha}
       + 2 C^{(t)}_{\alpha\beta} v_{(i)}^\beta \right)
       + 2 C^{(t)\beta\gamma} C^{(t)}_{\beta\gamma|\alpha}
       v_{(i)}^\alpha
       + 2 \left( \delta_{(i)} C^{(t)\alpha\beta} v_{(i)\beta}
       - 2 C^{(t)\alpha\beta} C^{(t)}_{\beta\gamma} v_{(i)}^\gamma
       \right)_{|\alpha} \right],
   \label{delta-i-eq-third} \\
   & & {1 \over a} \left[ a \left( 1 + \delta_{(i)} \right)
       v_{(i)\alpha} \right]^\cdot
       = \left( 1 + \delta_{(i)} \right) \kappa v_{(i)\alpha}
   \nonumber \\
   & & \qquad
       - {c \over a^2} \left\{
       \left( v_{(i)\beta} \chi^{,\beta} \right)_{,\alpha}
       \left( 1 + \delta_{(i)} \right)
       + \delta_{(i),\beta} \chi^{,\beta} v_{(i)\alpha}
       - 2 \left[ v_{(i)}^\beta
       \left( \chi_{,\beta} \varphi
       + \chi^{,\gamma} C^{(t)}_{\beta\gamma} \right)
       \right]_{,\alpha} \right\},
   \label{v-i-eq-third}
\eea where we recovered $c$. Equation (\ref{v-i-eq-third}) can be
written as \bea
   & & {1 \over a} \left( a v_{(i)\alpha} \right)^\cdot
       = - {c \over a^2} \left[ v_{(i)\beta} \chi^{,\beta}
       \left( 1 - 2 \varphi \right)
       - 2 v_{(i)}^\beta \chi^{,\gamma} C^{(t)}_{\beta\gamma}
       \right]_{,\alpha}.
   \label{v-i-eq-third-2}
\eea Equations (\ref{delta-eq-third}) and (\ref{kappa-eq-third}) are
the same as Eqs.\ (20) and (21) in \cite{third-order} which were
derived in the single component situation.

Combining Eqs.\ (\ref{delta-eq-third}),(\ref{kappa-eq-third}) and
Eqs.\ (\ref{kappa-eq-third})-(\ref{v-i-eq-third}), respectively, we
can derive \bea
   & & {1 \over a^2} \left[ a^2 \dot \delta
       + c \delta_{,\alpha} \chi^{,\alpha} \left( 1 - 2 \varphi
       \right)
       - 2 c \delta^{,\alpha} \chi^{,\beta} C^{(t)}_{\alpha\beta}
       \right]^\cdot
       - 4 \pi G \varrho \delta \left( 1 + \delta \right)
   \nonumber \\
   & & \qquad
       = {4 \over 3} \kappa^2 \left( 1 + \delta \right)
       - {c \over a^2} \left[ \kappa_{,\alpha} \chi^{,\alpha} \left(
       1 - 2 \varphi \right)
       + \left( \delta \kappa \right)_{,\alpha} \chi^{,\alpha}
       - 2 \kappa^{,\alpha} \chi^{,\beta} C^{(t)}_{\alpha\beta}
       \right]
   \nonumber \\
   & & \qquad
       + \left( {c \over a^2} \chi^{,\alpha|\beta}
       - {c \over 3} g^{(3)\alpha\beta} {\Delta \over a^2} \chi
       + \dot C^{(t)\alpha\beta} \right)
       \Bigg[
       \left( {c \over a^2} \chi_{,\alpha|\beta}
       + \dot C^{(t)}_{\alpha\beta} \right)
       \left( 1 - 4 \varphi + \delta \right)
   \nonumber \\
   & & \qquad
       - {4 c \over a^2} \chi_{,\alpha} \varphi_{,\beta}
       - {2 c \over a^2} \chi^{,\gamma} \left( 2
       C^{(t)}_{\gamma\alpha|\beta}
       - C^{(t)}_{\alpha\beta|\gamma} \right)
       - 4 C^{(t)}_{\beta\gamma}
       \left( {c \over a^2} \chi^{,\gamma}_{\;\;\;|\alpha}
       + \dot C^{(t)\gamma}_{\;\;\;\;\;\alpha} \right)
       - 4 \dot \varphi C^{(t)}_{\alpha\beta} \Bigg],
   \label{ddot-eq-CG-third} \\
   & & {1 \over a^2} \left[ a^2 \dot \delta_{(i)}
       + c \delta_{(i),\alpha} \chi^{,\alpha} \left( 1 - 2 \varphi
       \right)
       - 2 c \delta_{(i)}^{,\alpha} \chi^{,\beta} C^{(t)}_{\alpha\beta}
       \right]^\cdot
       - 4 \pi G \varrho \delta \left( 1 + \delta_{(i)} \right)
   \nonumber \\
   & & \qquad
       - {4 \over 3} \kappa^2 \left( 1 + \delta_{(i)} \right)
       + {c \over a^2} \left[ \kappa_{,\alpha} \chi^{,\alpha} \left(
       1 - 2 \varphi \right)
       + \left( \delta_{(i)} \kappa \right)_{,\alpha} \chi^{,\alpha}
       - 2 \kappa^{,\alpha} \chi^{,\beta} C^{(t)}_{\alpha\beta}
       \right]
   \nonumber \\
   & & \qquad
       - \left( {c \over a^2} \chi^{,\alpha|\beta}
       - {c \over 3} g^{(3)\alpha\beta} {\Delta \over a^2} \chi
       + \dot C^{(t)\alpha\beta} \right)
       \Bigg[
       \left( {c \over a^2} \chi_{,\alpha|\beta}
       + \dot C^{(t)}_{\alpha\beta} \right)
       \left( 1 - 4 \varphi + \delta_{(i)} \right)
   \nonumber \\
   & & \qquad
       - {4 c \over a^2} \chi_{,\alpha} \varphi_{,\beta}
       - {2 c \over a^2} \chi^{,\gamma} \left( 2
       C^{(t)}_{\gamma\alpha|\beta}
       - C^{(t)}_{\alpha\beta|\gamma} \right)
       - 4 C^{(t)}_{\beta\gamma}
       \left( {c \over a^2} \chi^{,\gamma}_{\;\;\;|\alpha}
       + \dot C^{(t)\gamma}_{\;\;\;\;\;\alpha} \right)
       - 4 \dot \varphi C^{(t)}_{\alpha\beta} \Bigg]
   \nonumber \\
   & & \qquad
       = {c \over a} \Bigg\{
       - 2 \left( \kappa - \dot \varphi \right)
       v^\alpha_{(i)|\alpha}
       - \left( \kappa + \dot \varphi \right)_{,\alpha}
       v^\alpha_{(i)}
       + c {\Delta \over a^2} \left( v_{(i)\alpha} \chi^{,\alpha}
       \right)
       + 2 v_{(i)}^{\alpha|\beta} \dot C^{(t)}_{\alpha\beta}
   \nonumber \\
   & & \qquad
       + v_{(i)}^\alpha \Big[ \left( 2 \varphi - \delta_{(i)}
       \right) \kappa_{,\alpha}
       - \left( \varphi + 2 \delta_{(i)} \right)_{,\alpha} \kappa
       + 2 \dot \varphi \delta_{(i),\alpha}
   \nonumber \\
   & & \qquad
       + \left( 4 \varphi \varphi_{,\alpha}
       - \varphi_{,\alpha} \delta_{(i)}
       - 2 \varphi^{,\beta} C^{(t)}_{\alpha\beta}
       + 2 C^{(t)\beta\gamma} C^{(t)}_{\beta\gamma|\alpha}
       + 2 \delta_{(i)}^{,\beta} C^{(t)}_{\alpha\beta}
       - 4 C^{(t)\beta\gamma} C^{(t)}_{\gamma\alpha|\beta}
       \right)^\cdot \Big]
   \nonumber \\
   & & \qquad
       + 2 v_{(i)}^{\alpha|\beta} \left[
       \kappa C^{(t)}_{\alpha\beta}
       + \left( - 4 \varphi C^{(t)}_{\alpha\beta}
       + \delta_{(i)} C^{(t)}_{\alpha\beta}
       - 2 C^{(t)\gamma}_{\;\;\;\;\;\beta} C^{(t)}_{\gamma\alpha}
       \right)^\cdot \right]
       + 2 v^\alpha_{(i)|\alpha}
       \left[ \kappa \left( 2 \varphi - \delta_{(i)} \right)
       + \dot \varphi \left( \delta_{(i)} - 4 \varphi \right) \right]
       \Bigg\}
   \nonumber \\
   & & \qquad
       + {c^2 \over a^3} \Bigg\{
       \left( \delta_{(i)} - 2 \varphi \right) \Delta \left(
       v_{(i)\alpha} \chi^{,\alpha} \right)
       + \left( v_{(i)\beta} \chi^{,\beta} \right)_{,\alpha}
       \delta_{(i)}^{,\alpha}
       + \left( \delta_{(i),\beta} \chi^{,\beta} v_{(i)\alpha}
       \right)^{|\alpha}
   \nonumber \\
   & & \qquad
       - 2 \Delta \left[ v_{(i)}^\alpha \left( \chi_{,\alpha}
       \varphi
       + \chi^{,\beta} C^{(t)}_{\alpha\beta} \right) \right]
       + \left( v_{(i)\beta} \chi^{,\beta} \right)^{,\alpha}
       \varphi_{,\alpha}
       - 2 \left( v_{(i)\gamma} \chi^{,\gamma}
       \right)^{,\alpha|\beta} C^{(t)}_{\alpha\beta} \Bigg\}.
   \label{ddot-eq-CG-i-third}
\eea All terms in the RHS of Eq.\ (\ref{ddot-eq-CG-i-third}) contain
$v_{(i)}^\alpha$ which decays to the linear order as \bea
   & & {\bf v}_{(i)} = {1 \over c} \left( {\bf u}_i - {\bf u}
       \right) \propto {1 \over a},
\eea see Eqs.\ (\ref{v_i}),(\ref{u_i-u}) and
\cite{second-order-multi}. By setting $\delta_{(i)} = \delta$ the
LHS of Eq.\ (\ref{ddot-eq-CG-i-third}) is the same as Eq.\
(\ref{ddot-eq-CG-third}). By setting $\delta_{(i)} = \delta$ and
$v^\alpha_{(i)} = 0$, Eq.\ (\ref{ddot-eq-CG-i-third}) becomes Eq.\
(\ref{ddot-eq-CG-third}).

In order to complete Eq.\ (\ref{ddot-eq-CG-i-third}) we need
equations for $\dot \varphi$, $v^\alpha_{(i)}$, and
$C^{(t)}_{\alpha\beta}$ terms to the second-order perturbations.
Equation for $v^\alpha_{(i)}$ is in Eq.\ (\ref{v-i-eq-third-2})
which gives \bea
   & & {1 \over a} \left( a v_{(i)\alpha} \right)^\cdot
       = - {c \over a^2} \left( v_{(i)\beta} \chi^{,\beta}
       \right)_{,\alpha},
\eea to the second order. The relation between $\kappa$ and $\chi$
can be found in Eq.\ (23) of \cite{third-order}. Recovering the
background curvature, from the momentum constraint equation in Eq.\
(\ref{Mom-constraint-ADM}), we can derive \bea
   & & \kappa + {c \over a^2} \left( \Delta + {R^{(3)} \over 2}
       \right) \chi
       = {c \over a^2} \left( 2 \varphi \Delta \chi - \chi^{,\alpha}
       \varphi_{,\alpha} \right)
       + C^{(t)\alpha\beta} \left(
       {2 c \over a^2} \chi_{,\alpha|\beta}
       - \dot C^{(t)}_{\alpha\beta}
       \right)
   \nonumber \\
   & & \qquad
       + {3 \over 2} \Delta^{-1} \nabla^\alpha \Bigg\{
       {c \over a^2} \left[ \chi^{,\beta} \varphi_{,\alpha|\beta}
       + \chi_{,\alpha} \left( \Delta + {2 \over 3} R^{(3)} \right)
       \varphi \right]
   \nonumber \\
   & & \qquad
       + \chi^{,\beta} {c \over a^2} \left( \Delta - {R^{(3)} \over
       3} \right) C^{(t)}_{\alpha\beta}
       - \varphi^{,\beta} \dot C^{(t)}_{\alpha\beta}
       + 2 C^{(t)\beta\gamma} \dot C^{(t)}_{\alpha\beta|\gamma}
       + C^{(t)}_{\beta\gamma|\alpha} \dot C^{(t)\beta\gamma}
       \Bigg\},
   \label{kappa-chi-eq}
\eea to the second order. Equation for $\varphi$ to the second order
follows from Eq.\ (99) in \cite{NL}. Ignoring ${\cal O}
(|v_{(i)\alpha}|^2)$, thus using Eq.\ (\ref{alpha-third}) for
$\alpha$ we can derive \bea
   & & 3 \dot \varphi
       = - \left( \kappa + c {\Delta \over a^2} \chi \right)
       + {c \over a^2} \left( 2 \varphi \Delta \chi - \chi^{,\alpha}
       \varphi_{,\alpha} \right)
       + 2 C^{(t)\alpha\beta} \left(
       {c \over a^2} \chi_{,\alpha|\beta}
       + \dot C^{(t)}_{\alpha\beta}
       \right).
   \label{dot-varphi-eq}
\eea In a flat background, the RHS is the second order; thus, to the
linear order we have $\dot \varphi = 0$. Equation for
$C^{(t)}_{\alpha\beta}$ to the second order is presented in Eq.\
(43) of \cite{third-order}.

\subsection{Pure general relativistic corrections}

Now, we {\it assume} \bea
   & & R^{(3)} = 0,
\eea thus $\dot \varphi = 0$ to the linear order. Based on the
apparent success in second-order perturbations, we continue to use
the {\it identifications} made in Eq.\ (\ref{identify-second}) valid
to the third order, thus \bea
   & & \kappa_v \equiv - {1 \over a} \nabla \cdot {\bf u}, \quad
       {\bf u} \equiv \nabla u, \quad
       {\bf v}_{(i)v} \equiv {1 \over c} \left( {\bf u}_i - {\bf u}
       \right), \quad
       \delta \equiv \delta_v, \quad
       \delta_i \equiv \delta_{(i)v}.
   \label{identify-third}
\eea In the following we consider pure scalar-type perturbations,
thus set $C^{(t)}_{\alpha\beta} \equiv 0$. Contributions from the
gravitational waves will be considered in the next section.

We need $\chi_v$ only to the second order. Equation
(\ref{kappa-chi-eq}) gives \bea
   & & \chi_v \equiv {a \over c} \left( u + \Delta^{-1} X \right),
\eea where \bea
   & & X \equiv 2 \varphi \nabla \cdot {\bf u}
       - {\bf u} \cdot \nabla \varphi
       + {3 \over 2} \Delta^{-1} \nabla \left[
       {\bf u} \cdot \nabla \left( \nabla \varphi \right)
       + {\bf u} \Delta \varphi \right].
   \label{X-eq-N-s}
\eea Equations (\ref{delta-eq-third})-(\ref{v-i-eq-third}) become
\bea
   & & \dot \delta
       + {1 \over a} \nabla \cdot {\bf u}
       = - {1 \over a} \nabla \cdot \left( \delta {\bf u} \right)
       + { 1\over a} \left( 2 \varphi {\bf u}
       - \nabla \Delta^{-1} X \right) \cdot \nabla \delta,
   \label{delta-eq-third-N-s} \\
   & & {1 \over a} \nabla \cdot \left( \dot {\bf u} + H {\bf u}
       \right)\
       + 4 \pi G \varrho \delta
       = - {1 \over a^2} \nabla \cdot \left( {\bf u} \cdot \nabla
       {\bf u} \right)
       - {\Delta \over a^2} \left[ {\bf u} \cdot \nabla \left(
       \Delta^{-1} X \right) \right]
       + {1 \over a^2} \left( {\bf u} \cdot \nabla X
       + {2 \over 3} X \nabla \cdot {\bf u} \right)
   \nonumber \\
   & & \qquad
       - {2 \over 3 a^2} \varphi {\bf u} \cdot \nabla \left( \nabla
       \cdot {\bf u} \right)
       + {4 \over a^2} \nabla \cdot \left[ \varphi \left(
       {\bf u} \cdot \nabla {\bf u}
       - {1 \over 3} {\bf u} \nabla \cdot {\bf u} \right) \right],
   \label{kappa-eq-third-N-s} \\
   & & \dot \delta_i + {1 \over a} \nabla \cdot {\bf u}_i
       = - {1 \over a} \nabla \cdot \left( \delta_i {\bf u}_i
       \right)
       + {1 \over a} \left[ 2 \varphi {\bf u}_i
       - \nabla \left( \Delta^{-1} X \right) \right] \cdot
       \nabla \delta_i
       + {1 \over a} \left[ 2 \varphi \nabla \cdot
       \left( {\bf u}_i - {\bf u} \right)
       - \left( {\bf u}_i - {\bf u} \right) \cdot \nabla \varphi \right]
   \nonumber \\
   & & \qquad
       + {2 \over a} \varphi \left[
       \delta_i \nabla \cdot \left( {\bf u}_i - {\bf u} \right)
       + 2 \left( {\bf u}_i - {\bf u} \right) \cdot \nabla \varphi
       - 2 \varphi \nabla \cdot \left( {\bf u}_i - {\bf u} \right)
       \right]
       - {1 \over a} \delta_i \left( {\bf u}_i - {\bf u} \right)
       \cdot \nabla \varphi,
   \label{delta-i-eq-third-N-s} \\
   & & {1 \over a} \nabla \cdot \left( \dot {\bf u}_i + H {\bf u}_i
       \right)
       + 4 \pi G \varrho \delta
       = - {1 \over a^2} \nabla \cdot \left( {\bf u}_i \cdot \nabla
       {\bf u}_i \right)
       - {\Delta \over a^2} \left[ {\bf u}_i \cdot \nabla \left(
       \Delta^{-1} X \right) \right]
       + {1 \over a^2} \left( {\bf u} \cdot \nabla X
       + {2 \over 3} X \nabla \cdot {\bf u} \right)
   \nonumber \\
   & & \qquad
       - {2 \over 3 a^2} \varphi {\bf u} \cdot \nabla \left( \nabla
       \cdot {\bf u} \right)
       + {4 \over a^2} \nabla \cdot \left[ \varphi \left(
       {\bf u} \cdot \nabla {\bf u}
       - {1 \over 3} {\bf u} \nabla \cdot {\bf u} \right) \right]
       + 2 {\Delta \over a^2} \left[ \varphi {\bf u} \cdot
       \left( {\bf u}_i - {\bf u} \right)
       \right].
   \label{v-i-eq-third-N-s}
\eea Equations (\ref{delta-eq-third-N}),(\ref{kappa-eq-third-N})
coincide with Eqs.\ (39),(40) in \cite{third-order}. Equation
(\ref{v-i-eq-third-2}) gives \bea
   & & {1 \over a} \left[ a \left( {\bf u}_i - {\bf u} \right)
       \right]^\cdot
       = - {1 \over a} \nabla \left\{
       \left( 1 - 2 \varphi \right) \left( {\bf u}_i - {\bf u}
       \right) \cdot \left[ {\bf u}
       + \nabla \left( \Delta^{-1} X \right) \right] \right\}.
   \label{v-i-eq-third-N-s-2}
\eea Equation (\ref{ddot-eq-CG-third}),(\ref{ddot-eq-CG-i-third})
give \bea
   & & {1 \over a^2} \left\{ a^2 \dot \delta
       + a \nabla \cdot \left( \delta {\bf u} \right)
       - a \left[ 2 \varphi {\bf u}
       - \nabla \left( \Delta^{-1} X \right) \right] \cdot \nabla
       \delta
       \right\}^\cdot
       - 4 \pi G \varrho \delta
       = {1 \over a^2} \nabla \cdot \left( {\bf u} \cdot \nabla
       {\bf u} \right)
       + {\Delta \over a^2} \left[ {\bf u} \cdot \nabla \left(
       \Delta^{-1} X \right) \right]
   \nonumber \\
   & & \qquad
       - {1 \over a^2} \left( {\bf u} \cdot \nabla X
       + {2 \over 3} X \nabla \cdot {\bf u} \right)
       + {2 \over 3 a^2} \varphi {\bf u} \cdot \nabla \left( \nabla
       \cdot {\bf u} \right)
       - {4 \over a^2} \nabla \cdot \left[ \varphi \left(
       {\bf u} \cdot \nabla {\bf u}
       - {1 \over 3} {\bf u} \nabla \cdot {\bf u} \right) \right],
   \label{ddot-eq-CG-third-N-s} \\
   & & {1 \over a^2} \left\{ a^2 \dot \delta_i
       + a \nabla \cdot \left( \delta_i {\bf u}_i \right)
       - a \left[ 2 \varphi {\bf u}_i
       - \nabla \left( \Delta^{-1} X \right) \right] \cdot \nabla
       \delta_i
       \right\}^\cdot
       - 4 \pi G \varrho \delta
       - {1 \over a^2} \nabla \cdot \left( {\bf u}_i \cdot \nabla
       {\bf u}_i \right)
       - {\Delta \over a^2} \left[ {\bf u}_i \cdot \nabla \left(
       \Delta^{-1} X \right) \right]
   \nonumber \\
   & & \qquad
       + {1 \over a^2} \left( {\bf u} \cdot \nabla X
       + {2 \over 3} X \nabla \cdot {\bf u} \right)
       - {2 \over 3 a^2} \varphi {\bf u} \cdot \nabla \left( \nabla
       \cdot {\bf u} \right)
       + {4 \over a^2} \nabla \cdot \left[ \varphi \left(
       {\bf u} \cdot \nabla {\bf u}
       - {1 \over 3} {\bf u} \nabla \cdot {\bf u} \right) \right]
   \nonumber \\
   & & \qquad
       = {1 \over a} \left[ 2 \dot \varphi \nabla \cdot
       \left( {\bf u}_i - {\bf u} \right)
       - \left( {\bf u}_i - {\bf u} \right) \cdot \nabla \dot \varphi
       \right]
       - 2 {\Delta \over a^2}
       \left[ \varphi {\bf u} \cdot \left( {\bf u}_i - {\bf u} \right) \right]
   \nonumber \\
   & & \qquad
       - {2 \over a^2} \varphi
       \left\{
       \Delta \left[ {\bf u} \cdot \left( {\bf u}_i - {\bf u}
       \right) \right]
       + \left( \nabla \cdot {\bf u}_i \right)
       \nabla \cdot \left( {\bf u}_i - {\bf u} \right) \right\}
       + {1 \over a^2} \left( \nabla \varphi \right) \cdot
       \left\{
       \nabla \left[ {\bf u} \cdot \left( {\bf u}_i - {\bf u}
       \right) \right]
       + \left( \nabla \cdot {\bf u}_i \right)
       \left( {\bf u}_i - {\bf u} \right) \right\}.
   \label{ddot-eq-CG-i-third-N-s}
\eea Equation (\ref{dot-varphi-eq}) becomes \bea
   & & \dot \varphi
       = {1 \over 3a} \left[ - X
       + 2 \varphi \nabla \cdot {\bf u}
       - {\bf u} \cdot \nabla \varphi \right]
       = - {1 \over 2} \Delta^{-1} \nabla \left[
       {1 \over a} {\bf u} \cdot \nabla \left( \nabla \varphi \right)
       + {\bf u} {\Delta \over a} \varphi \right].
   \label{dot-phi-eq-N-s}
\eea

A close examination of above equations reveals the followings.

(i) Equations
(\ref{delta-eq-third-N-s}),(\ref{kappa-eq-third-N-s}),(\ref{ddot-eq-CG-third-N-s})
are the same as Eqs.\ (25),(26),(28) in \cite{third-order} which are
valid in a single component fluid; we note, however, that ${\cal O}
(|v_{(i)\alpha}|^2)$ correction terms, which simply decay in an
expanding phase, appear even in the second-order perturbations, see
Eqs.\
(\ref{dot-delta-second}),(\ref{dot-u-second}),(\ref{ddot-delta-second})
above, or \cite{second-order-multi}.

(ii) If we ignore the $i$-indices, Eqs.
(\ref{delta-i-eq-third-N-s}),(\ref{v-i-eq-third-N-s}),(\ref{ddot-eq-CG-i-third-N-s})
are identical to Eqs.\
(\ref{delta-eq-third-N-s}),(\ref{kappa-eq-third-N-s}),(\ref{ddot-eq-CG-third-N-s}),
respectively; we note, however, that the presence of ${\cal O}
(|v_{(i)\alpha}|^2)$ correction terms causes differences even in the
second-order perturbations, see Eqs.\
(\ref{E-conservation-zero-pressure-irrotation-total-CG})-(\ref{ddot-eq-CG-i}).

(iii) We already showed that, to the second order, even in the case
of multi-component, the general relativistic equations are identical
to the Newtonian ones, thus having relativistic/Newtonian
correspondence. The presence of ${\cal O} (|v_{(i)\alpha}|^2)$
correction terms may cause differences, but we have shown that these
corrections are simply decaying in the expanding phase.

(iv) The pure third-order correction terms in the above equations
all involve $\varphi$-term to the linear order in various forms of
convolution with the second-order terms. As we took the comoving
gauge $v \equiv 0$, the spatial curvature variable $\varphi$ is the
same as a gauge-invariant combination $\varphi_v \equiv \varphi - a
H v/c$ to the linear order.

(v) We note that, to the linear order, $\varphi_v$ is one of the
well known conserved quantity in the large-scale
\cite{H-PRW-1991,conserved}. For $K = 0$, but considering general
$\Lambda$, we have $\dot \varphi_v = 0$, thus \bea
   & & \varphi_v = C({\bf x}),
\eea with vanishing decaying mode (in an expanding phase) to the
leading order in the large-scale expansion
\cite{Hwang-Noh-Newtonian}.

(vi) We also note that the value of $\varphi_v$ in the large-scale
is of the same order as the gravitational potential fluctuation
$\varphi_\chi$ which is again of the same order as the relative
temperature fluctuations $\delta T/T$ of the cosmic microwave
background radiation (CMB). In general we have \cite{third-order}
\bea
   & & \varphi_v = \varphi_\chi - a H v_\chi/c
       = - \delta \Phi/c^2
       + \dot a \Delta^{-1} \nabla \cdot {\bf u}/c^2,
   \label{varphi_v}
\eea to the linear order, where we have $\varphi_\chi = - \delta
\Phi/c^2$ and ${\bf u} = - c \nabla v_\chi$; $\delta \Phi$ is the
Newtonian gravitational potential identified to the linear order in
\cite{Harrison-1967,Bardeen-1980,Hwang-Noh-Newtonian}. For $K = 0 =
\Lambda$ we have \bea
   & & \varphi_v = {5 \over 3} \varphi_\chi.
\eea The temperature anisotropy of CMB, in a flat background without
the cosmological constant, gives \cite{SW-1967,HN-SW1} \bea
   & & {\delta T \over T} \sim {1 \over 3} \varphi_\chi
       = {1 \over 3} {\delta \Phi \over c^2}
       \sim {1 \over 5} \varphi_v \sim {1 \over 5} C,
   \label{SW}
\eea to the linear order; this is a part of the Sachs-Wolfe effect
\cite{SW-1967}. The COBE observations of CMB give ${\delta T / T}
\sim 10^{-5}$ \cite{CMB}, thus \bea
   & & \varphi_v \sim 5 \times 10^{-5},
   \label{CMB-constraint}
\eea in the large-scale limit near horizon scale. Therefore, the
pure general relativistic third-order correction terms are
independent of the presence of the horizon scale and are smaller by
factor $5 \times 10^{-5}$ compared with the second-order
relativistic/Newtonian terms due to the low level anisotropies of
the cosmic microwave background radiation.

\subsection{Contributions from tensor-type perturbations}

Here we present set of equations describing the scalar-type
perturbation equations to the third order, now including the
contributions from the tensor-type perturbations. We continue to use
the Newtonian variables identified in Eq.\ (\ref{identify-third}).

Equations (\ref{delta-eq-third})-(\ref{v-i-eq-third}) become \bea
   & & \dot \delta
       + {1 \over a} \nabla \cdot {\bf u}
       = - {1 \over a} \nabla \cdot \left( \delta {\bf u} \right)
       + { 1\over a} \left( 2 \varphi {\bf u}
       - \nabla \Delta^{-1} X \right) \cdot \nabla \delta
       + {2 \over a} \delta^{,\alpha} u^\beta C^{(t)}_{\alpha\beta},
   \label{delta-eq-third-N} \\
   & & {1 \over a} \nabla \cdot \left( \dot {\bf u} + H {\bf u}
       \right)\
       + 4 \pi G \varrho \delta
       = - {1 \over a^2} \nabla \cdot \left( {\bf u} \cdot \nabla
       {\bf u} \right)
       - \dot C^{(t)\alpha\beta} \left( {2 \over a} u_{\alpha|\beta}
       + \dot C^{(t)}_{\alpha\beta} \right)
   \nonumber \\
   & & \qquad
       - {\Delta \over a^2} \left[ {\bf u} \cdot \nabla \left(
       \Delta^{-1} X \right) \right]
       + {1 \over a^2} \left( {\bf u} \cdot \nabla X
       + {2 \over 3} X \nabla \cdot {\bf u} \right)
       - {2 \over 3 a^2} \varphi {\bf u} \cdot \nabla \left( \nabla
       \cdot {\bf u} \right)
       + {4 \over a^2} \nabla \cdot \left[ \varphi \left(
       {\bf u} \cdot \nabla {\bf u}
       - {1 \over 3} {\bf u} \nabla \cdot {\bf u} \right) \right]
   \nonumber \\
   & & \qquad
       + {2 \over a^2} u^\alpha \nabla^\beta \left( \nabla \cdot
       {\bf u} \right) C^{(t)}_{\alpha\beta}
       + 2 \dot C^{(t)\alpha\beta} \left[
       {4 \over a} \varphi u_{\alpha|\beta}
       + {2 \over a} u_\alpha \nabla_\beta \varphi
       + 2 \varphi \dot C^{(t)}_{\alpha\beta}
       - {1 \over a} \left( \Delta^{-1} X \right)_{,\alpha|\beta}
       \right]
   \nonumber \\
   & & \qquad
       + 2 \left( {1 \over a} u^{\alpha|\beta}
       + \dot C^{(t)\alpha\beta} \right)
       \left[ - {2 \over 3 a} \left( \nabla \cdot {\bf u} \right)
       C^{(t)}_{\alpha\beta}
       + {1 \over a} u^\gamma \left( 2 C^{(t)}_{\gamma\alpha|\beta}
       - C^{(t)}_{\alpha\beta|\gamma} \right)
       + 2 C^{(t)}_{\beta\gamma} \left( {1 \over a}
       u^{\gamma}_{\;\;|\alpha}
       + \dot C^{(t)\gamma}_{\;\;\;\;\;\alpha} \right) \right],
   \label{kappa-eq-third-N} \\
   & & \dot \delta_i + {1 \over a} \nabla \cdot {\bf u}_i
       = - {1 \over a} \nabla \cdot \left( \delta_i {\bf u}_i
       \right)
       + {1 \over a} \left[ 2 \varphi {\bf u}_i
       - \nabla \left( \Delta^{-1} X \right) \right] \cdot
       \nabla \delta_i
       + {2 \over a} \delta_i^{,\alpha} u_i^\beta
       C^{(t)}_{\alpha\beta}
   \nonumber \\
   & & \qquad
       + {1 \over a} \left[ 2 \varphi \nabla \cdot
       \left( {\bf u}_i - {\bf u} \right)
       - \left( {\bf u}_i - {\bf u} \right) \cdot \nabla \varphi
       + 2 C^{(t)}_{\alpha\beta} \left( u_i^\alpha - u^\alpha
       \right)^{|\beta} \right]
   \nonumber \\
   & & \qquad
       + {2 \over a} \varphi \left[
       \delta_i \nabla \cdot \left( {\bf u}_i - {\bf u} \right)
       + 2 \left( {\bf u}_i - {\bf u} \right) \cdot \nabla \varphi
       - 2 \varphi \nabla \cdot \left( {\bf u}_i - {\bf u} \right)
       \right]
       - {1 \over a} \delta_i \left( {\bf u}_i - {\bf u} \right)
       \cdot \nabla \varphi
   \nonumber \\
   & & \qquad
       + {2 \over a} C^{(t)\alpha\beta} \left\{
       \left( \delta_i - 4 \varphi \right)
       \left( u_{i\alpha} - u_\alpha \right)_{|\beta}
       - \varphi_{,\alpha} \left( u_{i\beta} - u_\beta \right)
       + C^{(t)}_{\alpha\beta|\gamma}
       \left( u_i^\gamma - u^\gamma \right)
       - 2 \left[ C^{(t)}_{\beta\gamma}
       \left( u_i^\gamma - u^\gamma \right) \right]_{|\alpha}
       \right\},
   \label{delta-i-eq-third-N} \\
   & & {1 \over a} \nabla \cdot \left( \dot {\bf u}_i + H {\bf u}_i
       \right)
       + 4 \pi G \varrho \delta
       = - {1 \over a^2} \nabla \cdot \left( {\bf u}_i \cdot \nabla
       {\bf u}_i \right)
       - \dot C^{(t)\alpha\beta} \left( {2 \over a} u_{\alpha|\beta}
       + \dot C^{(t)}_{\alpha\beta} \right)
   \nonumber \\
   & & \qquad
       - {\Delta \over a^2} \left[ {\bf u}_i \cdot \nabla \left(
       \Delta^{-1} X \right) \right]
       + {1 \over a^2} \left( {\bf u} \cdot \nabla X
       + {2 \over 3} X \nabla \cdot {\bf u} \right)
   \nonumber \\
   & & \qquad
       - {2 \over 3 a^2} \varphi {\bf u} \cdot \nabla \left( \nabla
       \cdot {\bf u} \right)
       + {4 \over a^2} \nabla \cdot \left[ \varphi \left(
       {\bf u} \cdot \nabla {\bf u}
       - {1 \over 3} {\bf u} \nabla \cdot {\bf u} \right) \right]
       + 2 {\Delta \over a^2} \left[ \varphi {\bf u} \cdot
       \left( {\bf u}_i - {\bf u} \right)
       + u^\alpha
       \left( u_i^\beta - u^\beta \right) C^{(t)}_{\alpha\beta}
       \right]
   \nonumber \\
   & & \qquad
       + {2 \over a^2} u^\alpha \nabla^\beta \left( \nabla \cdot
       {\bf u} \right) C^{(t)}_{\alpha\beta}
       + 2 \dot C^{(t)\alpha\beta} \left[
       {4 \over a} \varphi u_{\alpha|\beta}
       + {2 \over a} u_\alpha \nabla_\beta \varphi
       + 2 \varphi \dot C^{(t)}_{\alpha\beta}
       - {1 \over a} \left( \Delta^{-1} X \right)_{,\alpha|\beta}
       \right]
   \nonumber \\
   & & \qquad
       + 2 \left( {1 \over a} u^{\alpha|\beta}
       + \dot C^{(t)\alpha\beta} \right)
       \left[ - {2 \over 3 a} \left( \nabla \cdot {\bf u} \right)
       C^{(t)}_{\alpha\beta}
       + {1 \over a} u^\gamma \left( 2 C^{(t)}_{\gamma\alpha|\beta}
       - C^{(t)}_{\alpha\beta|\gamma} \right)
       + 2 C^{(t)}_{\beta\gamma} \left( {1 \over a}
       u^{|\gamma}_{\;\;\;\alpha}
       + \dot C^{(t)\gamma}_{\;\;\;\;\;\alpha} \right) \right].
   \label{v-i-eq-third-N}
\eea Equations (\ref{delta-eq-third-N}),(\ref{kappa-eq-third-N})
coincide with Eqs.\ (39),(40) in \cite{third-order}. Equation
(\ref{v-i-eq-third-2}) gives \bea
   & & {1 \over a} \left[ a \left( {\bf u}_i - {\bf u} \right)
       \right]^\cdot
       = - {1 \over a} \nabla \left\{
       \left( 1 - 2 \varphi \right) \left( {\bf u}_i - {\bf u}
       \right) \cdot \left[ {\bf u}
       + \nabla \left( \Delta^{-1} X \right) \right]
       - 2 u^\alpha \left( u_i^\beta - u^\beta \right)
       C^{(t)}_{\alpha\beta} \right\}.
   \label{v-i-eq-third-N-2}
\eea Equation (\ref{v-i-eq-third-2}) gives \bea
   & & {1 \over a^2} \left\{ a^2 \dot \delta
       + a \nabla \cdot \left( \delta {\bf u} \right)
       - a \left[ 2 \varphi {\bf u}
       - \nabla \left( \Delta^{-1} X \right) \right] \cdot \nabla
       \delta
       - 2 a \delta^{,\alpha} u^\beta C^{(t)}_{\alpha\beta}
       \right\}^\cdot
       - 4 \pi G \varrho \delta
   \nonumber \\
   & & \qquad
       = {1 \over a^2} \nabla \cdot \left( {\bf u} \cdot \nabla
       {\bf u} \right)
       + \dot C^{(t)\alpha\beta} \left( {2 \over a} u_{\alpha|\beta}
       + \dot C^{(t)}_{\alpha\beta} \right)
       + {\Delta \over a^2} \left[ {\bf u} \cdot \nabla \left(
       \Delta^{-1} X \right) \right]
       - {1 \over a^2} \left( {\bf u} \cdot \nabla X
       + {2 \over 3} X \nabla \cdot {\bf u} \right)
   \nonumber \\
   & & \qquad
       + {2 \over 3 a^2} \varphi {\bf u} \cdot \nabla \left( \nabla
       \cdot {\bf u} \right)
       - {4 \over a^2} \nabla \cdot \left[ \varphi \left(
       {\bf u} \cdot \nabla {\bf u}
       - {1 \over 3} {\bf u} \nabla \cdot {\bf u} \right) \right]
   \nonumber \\
   & & \qquad
       - {2 \over a^2} u^\alpha \nabla^\beta \left( \nabla \cdot
       {\bf u} \right) C^{(t)}_{\alpha\beta}
       - 2 \dot C^{(t)\alpha\beta} \left[
       {4 \over a} \varphi u_{\alpha|\beta}
       + {2 \over a} u_\alpha \nabla_\beta \varphi
       + 2 \varphi \dot C^{(t)}_{\alpha\beta}
       - {1 \over a} \left( \Delta^{-1} X \right)_{,\alpha|\beta}
       \right]
   \nonumber \\
   & & \qquad
       - 2 \left( {1 \over a} u^{\alpha|\beta}
       + \dot C^{(t)\alpha\beta} \right)
       \left[ - {2 \over 3 a} \left( \nabla \cdot {\bf u} \right)
       C^{(t)}_{\alpha\beta}
       + {1 \over a} u^\gamma \left( 2 C^{(t)}_{\gamma\alpha|\beta}
       - C^{(t)}_{\alpha\beta|\gamma} \right)
       + 2 C^{(t)}_{\beta\gamma} \left( {1 \over a}
       u^{\gamma}_{\;\;|\alpha}
       + \dot C^{(t)\gamma}_{\;\;\;\;\;\alpha} \right) \right],
   \label{ddot-eq-CG-third-N} \\
   & & {1 \over a^2} \left\{ a^2 \dot \delta_i
       + a \nabla \cdot \left( \delta_i {\bf u}_i \right)
       - a \left[ 2 \varphi {\bf u}_i
       - \nabla \left( \Delta^{-1} X \right) \right] \cdot \nabla
       \delta_i
       - 2 a \delta_i^{,\alpha} u_i^\beta C^{(t)}_{\alpha\beta}
       \right\}^\cdot
       - 4 \pi G \varrho \delta
   \nonumber \\
   & & \qquad
       - {1 \over a^2} \nabla \cdot \left( {\bf u}_i \cdot \nabla
       {\bf u}_i \right)
       - \dot C^{(t)\alpha\beta} \left( {2 \over a} u_{i\alpha|\beta}
       + \dot C^{(t)}_{\alpha\beta} \right)
       - {\Delta \over a^2} \left[ {\bf u}_i \cdot \nabla \left(
       \Delta^{-1} X \right) \right]
       + {1 \over a^2} \left( {\bf u} \cdot \nabla X
       + {2 \over 3} X \nabla \cdot {\bf u} \right)
   \nonumber \\
   & & \qquad
       - {2 \over 3 a^2} \varphi {\bf u} \cdot \nabla \left( \nabla
       \cdot {\bf u} \right)
       + {4 \over a^2} \nabla \cdot \left[ \varphi \left(
       {\bf u} \cdot \nabla {\bf u}
       - {1 \over 3} {\bf u} \nabla \cdot {\bf u} \right) \right]
   \nonumber \\
   & & \qquad
       + {2 \over a^2} u^\alpha \nabla^\beta \left( \nabla \cdot
       {\bf u} \right) C^{(t)}_{\alpha\beta}
       + 2 \dot C^{(t)\alpha\beta} \left[
       {4 \over a} \varphi u_{\alpha|\beta}
       + {2 \over a} u_\alpha \nabla_\beta \varphi
       + 2 \varphi \dot C^{(t)}_{\alpha\beta}
       - {1 \over a} \left( \Delta^{-1} X \right)_{,\alpha|\beta}
       \right]
   \nonumber \\
   & & \qquad
       + 2 \left( {1 \over a} u^{\alpha|\beta}
       + \dot C^{(t)\alpha\beta} \right)
       \left[ - {2 \over 3 a} \left( \nabla \cdot {\bf u} \right)
       C^{(t)}_{\alpha\beta}
       + {1 \over a} u^\gamma \left( 2 C^{(t)}_{\gamma\alpha|\beta}
       - C^{(t)}_{\alpha\beta|\gamma} \right)
       + 2 C^{(t)}_{\beta\gamma} \left( {1 \over a}
       u^{|\gamma}_{\;\;\;\alpha}
       + \dot C^{(t)\gamma}_{\;\;\;\;\;\alpha} \right) \right]
   \nonumber \\
   & & \qquad
       = {1 \over a} \left[ 2 \dot \varphi \nabla \cdot
       \left( {\bf u}_i - {\bf u} \right)
       - \left( {\bf u}_i - {\bf u} \right) \cdot \nabla \dot \varphi
       \right]
       - 2 {\Delta \over a^2}
       \left[ \varphi {\bf u} \cdot \left( {\bf u}_i - {\bf u} \right)
       + u^\alpha \left( u_i^\beta - u^\beta \right)
       C^{(t)}_{\alpha\beta} \right]
   \nonumber \\
   & & \qquad
       - {2 \over a^2} \varphi
       \left\{
       \Delta \left[ {\bf u} \cdot \left( {\bf u}_i - {\bf u}
       \right) \right]
       + \left( \nabla \cdot {\bf u}_i \right)
       \nabla \cdot \left( {\bf u}_i - {\bf u} \right) \right\}
       + {1 \over a^2} \left( \nabla \varphi \right) \cdot
       \left\{
       \nabla \left[ {\bf u} \cdot \left( {\bf u}_i - {\bf u}
       \right) \right]
       + \left( \nabla \cdot {\bf u}_i \right)
       \left( {\bf u}_i - {\bf u} \right) \right\}
   \nonumber \\
   & & \qquad
       - {2 \over a^2} C^{(t)\alpha\beta}
       \left\{
       \left[ {\bf u} \cdot \left( {\bf u}_i - {\bf u}
       \right) \right]_{,\alpha|\beta}
       + \left( \nabla \cdot {\bf u}_i \right)
       \left( u_{i\alpha} - u_\alpha \right)_{|\beta}
       - a \dot C^{(t)}_{\alpha\beta|\gamma}
       \left( u_i^\gamma - u^\gamma \right)
       + 2 a \left[ \dot C^{(t)}_{\beta\gamma}
       \left( u_i^\gamma - u^\gamma \right) \right]_{|\alpha}
       \right\}
   \nonumber \\
   & & \qquad
       + {2 \over a} \dot C^{(t)\alpha\beta} \left\{
       \left( \delta_i - 4 \varphi \right)
       \left( u_{i\alpha} - u_\alpha \right)_{|\beta}
       - \varphi_{,\alpha} \left( u_{i\beta} - u_\beta \right)
       + C^{(t)}_{\alpha\beta|\gamma}
       \left( u_i^\gamma - u^\gamma \right)
       - 2 \left[ C^{(t)}_{\beta\gamma}
       \left( u_i^\gamma - u^\gamma \right) \right]_{|\alpha}
       \right\}.
   \label{ddot-eq-CG-i-third-N}
\eea Equations (\ref{kappa-chi-eq}),(\ref{dot-varphi-eq}) become
\bea
   & & X \equiv 2 \varphi \nabla \cdot {\bf u}
       - {\bf u} \cdot \nabla \varphi
       + C^{(t)\alpha\beta} \left(
       2 u_{\alpha|\beta} - a \dot C^{(t)}_{\alpha\beta} \right)
   \nonumber \\
   & & \qquad
       + {3 \over 2} \Delta^{-1} \nabla^\alpha \left[
       {\bf u} \cdot \nabla \left( \nabla_\alpha \varphi \right)
       + u_\alpha \Delta \varphi
       + u^\beta \Delta C^{(t)}_{\alpha\beta}
       - a \varphi^{,\beta} \dot C^{(t)}_{\alpha\beta}
       + 2 a C^{(t)\beta\gamma} \dot C^{(t)}_{\alpha\beta|\gamma}
       + a C^{t)}_{\beta\gamma|\alpha}
       \dot C^{(t)\beta\gamma} \right],
   \label{X-eq-N} \\
   & & \dot \varphi
       = {1 \over 3a} \left[ - X
       + 2 \varphi \nabla \cdot {\bf u}
       - {\bf u} \cdot \nabla \varphi
       + 2 C^{(t)\alpha\beta}
       \left( u_{\alpha|\beta}
       + a \dot C^{(t)}_{\alpha\beta} \right) \right]
   \nonumber \\
   & & \quad
       = C^{(t)\alpha\beta} \dot C^{(t)}_{\alpha\beta}
       - {1 \over 2} \Delta^{-1} \nabla^\alpha \left[
       {1 \over a} {\bf u} \cdot \nabla \left( \nabla_\alpha \varphi \right)
       + u_\alpha {\Delta \over a} \varphi
       + u^\beta {\Delta \over a} C^{(t)}_{\alpha\beta}
       - \varphi^{,\beta} \dot C^{(t)}_{\alpha\beta}
       + 2 C^{(t)\beta\gamma} \dot C^{(t)}_{\alpha\beta|\gamma}
       + C^{t)}_{\beta\gamma|\alpha}
       \dot C^{(t)\beta\gamma} \right].
   \label{dot-phi-eq-N}
\eea

%
%
\section{Discussion}
                                      \label{sec:Discussion}

In this work we have successfully derived pure general relativistic
correction terms appearing in the third-order perturbations of the
zero-pressure irrotational multi-component fluids in a flat
background. We have ignored ${\cal O} (|{\bf u} - {\bf u}_i|^2)$
correction terms which simply decay in an expanding phase. Our main
results are presented in Eqs.\
(\ref{identify-third})-(\ref{dot-phi-eq-N-s}), and in Eqs.\
(\ref{delta-eq-third-N})-(\ref{dot-phi-eq-N}) in the presence of the
gravitational waves. The equations for the collective component are
identical to the ones in the single component case. If we further
ignore ${\cal O} ({\bf u} - {\bf u}_i)$ correction terms, which are
again simply decaying in an expanding phase, the forms of
relativistic correction terms in the individual component are the
same as the ones in the collective component. Our results show that,
even in the multi-component situation, the pure relativistic
correction terms are smaller by a factor \bea
   & & \varphi_v \sim {5 \over 3} {\delta \Phi \over c^2}
       \sim 5 {\delta T \over T} \sim 5 \times 10^{-5},
\eea compared with the relativistic/Newtonian second-order terms and
are independent of the presence of horizon scale.

By taking different temporal gauge condition (hypersurface
condition) we can easily introduce apparently horizon dependent
terms with arbitrarily huge amplitudes. The exact
relativistic/Newtonian correspondence to the second-order
perturbations was available essentially due to our proper choice of
gauge conditions and correct identifications of relativistic
variables with the Newtonian ones. In our third-order extension we
have assumed the same identification holds even in the third-order
perturbations, which might not be necessarily the unique choice.
However, the properties (the smallness of the amplitudes and
independence from the horizon scale) of our third-order correction
terms assure that our choice of the gauge and identifications are
very likely to be the correct and best ones even to the third order.

Our results may have practically important implications in currently
favored cosmological pursuits by assuring the use of Newtonian
physics in the large-scale nonlinear processes which often involve
two-component zero-pressure fluids (say, dust and cold dark matter).
As we have shown the exact relativistic/Newtonian correspondence to
the second order and small horizon-independent third-order
correction terms, it is now more secure to use Newtonian physics
near (and even beyond) the horizon scale which is indeed a
noticeable trend in current cosmological simulations
\cite{simulation}. At this point it might be also worth emphasizing
the effects of rotational perturbations. In
\cite{second-order-multi} we have shown that the rotational
perturbations to the second-order also have relativistic/Newtonian
correspondence in the small-scale (sub-horizon scale) limit. As the
numerical simulations naturally include the rotational mode, this
might be another good news to the cosmology community based on
Newtonian physics.

Although we have estimated that the third-order pure general
relativistic correction terms are small in the current large-scale
structures, it would be still interesting to see whether cosmology
could reach a stage where $\varphi_v \sim 5 \times 10^{-5}$-factor
smaller correction terms could have noticeable effect on the
large-scale structure formation processes. In either case the
third-order terms we have presented in this work are the first
non-vanishing general relativistic correction terms in Newtonian
nonlinear equations, and may have historical as well as practical
importance in cosmology. Analytic studies in the Newtonian context
of a single component zero-pressure, irrotational fluid have been
actively investigated in \cite{quasilinear}. Whether our general
relativistic corrections appearing in the third order (both in the
single-component and multi-component cases), and corrections
appearing even in the second order by effects of pressure, rotation,
and background curvature, will have noticeable roles in the
large-scale structure formation process are left for future
investigations.

%
%
\subsection*{Acknowledgments}

H.N. was supported by grant No.\ C00022 from the Korea Research
Foundation.

%
%


\begin{thebibliography}{}
\bibitem{NL}
         H. Noh and J. Hwang, Phys. Rev. D \textbf{69}, 104011 (2004),
                      Preprint astro-ph/0305123.
\bibitem{second-order}
         H. Noh and J. Hwang, Class. Quant. Grav. \textbf{22}, 3181 (2005),
                      Preprint gr-qc/0412127;
         J. Hwang and H. Noh, Phys. Rev. D \textbf{72}, 044011 (2005),
                      Preprint gr-qc/0412128;
         J. Hwang and H. Noh, Monthly Not. R. Astron. Soc. \textbf{367}, 1515 (2006),
                      Preprint astro-ph/0507159;
         H. Noh and J. Hwang, Gen. Rel. Grav. \textbf{38}, 703 (2006),
                      Preprint astro-ph/0512636.
\bibitem{second-order-SG}
         J. Hwang and H. Noh, Phys. Rev. D \textbf{73}, 044021 (2006),
                      Preprint astro-ph/0601041.
\bibitem{third-order}
         J. Hwang and H. Noh, Phys. Rev. D \textbf{72}, 044012 (2005),
                      Preprint gr-qc/0412129.
\bibitem{second-order-multi}
         J. Hwang, and H. Noh, Phys. Rev. D, submitted,
                      Preprint astro-ph/0704.1927.
\bibitem{Friedmann-1922}
         A.A. Friedmann, Zeitschrift f\"ur Physik {\bf 10}, 377 (1922),
                         and
                         {\it ibid.} {\bf 21}, 326 (1924);
                         both papers are translated in
                         {\it Cosmological-constants: papers in modern
                         cosmology}, edited by J. Bernstein and G. Feinberg
                         (Columbia Univ. Press, New York, 1986), p49 and p59;
         H.P. Robertson, Proceedings of the National Academy of Science
                         {\bf 15}, 822 (1929).
\bibitem{Milne-1934}
         E.A. Milne, Quart. J. Math. {\bf 5}, 64 (1934);
         W.H. McCrea and E.A. Milne, Quart. J. Math. {\bf 5}, 73 (1934).
\bibitem{Layzer-1954}
         D. Layzer, Astron. J. \textbf{59}, 268 (1954);
         D.S. Lemons, Am. J. Phys. \textbf{56}, 502 (1988).
\bibitem{Lifshitz-1946}
         E.M. Lifshitz, J. Phys. (USSR) \textbf{10}, 116 (1946);
         E.M. Lifshitz and I.M. Khalatnikov, Adv. Phys. {\bf 12}, 185 (1963).
\bibitem{Bonnor-1957}
         W.B. Bonnor, Mon. Not. R. Astron. Soc. \textbf{117}, 104 (1957).
\bibitem{Peebles-1980}
         P.J.E. Peebles, \emph{The large-scale structure of the universe},
                         (Princeton Univ. Press, Princeton, 1980).
\bibitem{covariant}
         J. Ehlers, Proceedings of the mathematical-natural science of
                    the Mainz academy of science and literature,
                    Nr. {\bf 11}, 792 (1961),
                    translated in Gen. Rel. Grav. {\bf 25}, 1225 (1993);
         G.F.R. Ellis, in {\it General relativity and cosmology,
                       Proceedings of
                       the international summer school of physics Enrico
                       Fermi course 47}, edited by R. K. Sachs (Academic
                       Press, New York, 1971), p104;
         in {\it Cargese Lectures in Physics}, edited by
                       E. Schatzmann (Gorden and Breach, New York, 1973), p1.
\bibitem{ADM}
         R. Arnowitt, S. Deser, and C.W. Misner, in {\it Gravitation: an
                      introduction to current research}, edited by  L. Witten
                      (Wiley, New York, 1962) p. 227.
\bibitem{Bardeen-1980}
         J.M. Bardeen, Phys. Rev. D \textbf{22}, 1882 (1980).
\bibitem{HV-1990}
         J. Hwang and E.T. Vishniac, Astrophys. J. {\bf 353}, 1
                  (1990).
\bibitem{Bardeen-1988}
         J.M. Bardeen, \emph{Particle Physics and Cosmology},
              edited by L. Fang and A. Zee (Gordon and Breach, London,
              1988) p1.
\bibitem{HN-multi-CQG}
         J. Hwang and H. Noh, Class. Quant. Grav. {\bf 19}, 527 (2002)
                  Preprint astro-ph/0103244.
\bibitem{Hwang-Noh-Newtonian}
         J. Hwang and H. Noh, Gen. Rel. Grav. {\bf 31}, 1131 (1999),
                  Preprint astro-ph/9907063.
\bibitem{Nariai-1969}
         H. Nariai, Prog. Theor. Phys. {\bf 41}, 686 (1969).
\bibitem{H-PRW-1991}
         J. Hwang, Astrophys. J. {\bf 375}, 443 (1991).
\bibitem{conserved}
         J. Hwang and H. Noh, Phys. Rev. D {\bf 71}, 063536 (2005),
                  Preprint gr-qc/0412126 .
\bibitem{Harrison-1967}
         E.R. Harrison, Rev. Mod. Phys. {\bf 39}, 862 (1967).
\bibitem{SW-1967}
         R.K. Sachs and A.M. Wolfe, Astrophys. J. {\bf 147}, 73 (1967).
\bibitem{HN-SW1}
         J. Hwang and H. Noh, Phys. Rev. D {\bf 59}, 067302 (1999),
                   Preprint astro-ph/9812007.
\bibitem{CMB}
         G.F. Smoot, {\it et al.} Astrophys. J. {\bf 396}, L1 (1992).
\bibitem{simulation}
         J.M. Colberg, et al., Monthly Not. R. Astron. Soc. {\bf 319}, 209 (2000),
                 Preprint astro-ph/0005259;
         A. Jenkins, et al., Monthly Not. R. Astron. Soc. {\bf 321}, 372 (2001),
                 Preprint astro-ph/0005260;
         A.E. Evrard, et al., Astrophys. J. {\bf 573}, 7 (2002),
                 Preprint astro-ph/0110246;
         P. Bode and J.P. Ostriker, Astrophys. J. Suppl. {\bf 145}, 1 (2003),
                 Preprint astro-ph/0302065;
         J. Dubinski, J. Kim, C. Park R. Humble, New Astron., {\bf 9}, 111
                 (2003), Preprint astro-ph/0304467;
         V. Springel, et al., Nature {\bf 435}, 629 (2005),
                 Preprint astro-ph/0504097;
         V. Springel, C.S. Frenk, S.D.M. White, Nature {\bf 440}, 1137
                 (2006), Preprint astro-ph/0604561.
\bibitem{quasilinear}
         E.T. Vishniac, Monthly Not. R. Astron. Soc. {\bf 203}, 345 (1983);
         M.H. Goroff, B. Grinstein, S.-J. Rey, and M.B. Wise,
                        Astrophys. J., {\bf 311}, 6 (1986);
         F. Bernardeau, S. Colombi, E. Gaztanaga, and R. Scoccimarro,
                        Phys. Rep. {\bf 367}, 1 (2002),
                        Preprint astro-ph/0112551.
\end{thebibliography}
\end{document}